\documentclass[aps,twocolumn,prd,showpacs,nofootinbib,floatfix]{revtex4}
\usepackage{graphicx}
\usepackage{amsmath}
\usepackage{amssymb}

\newcommand{\ie}{i.e.~}
\newcommand{\eg}{e.g.~}

\newcommand{\diag}{\negmedspace \mathrm{diag}}

\newcommand{\jordan}[1]{\tilde{#1}}

\newcommand{\sss}[1]{{\scriptscriptstyle{#1}}}
\newcommand{\vsep}{\vspace{2pt}}
\newcommand{\zero}{{\sss{0}}}
\newcommand{\one}{{\sss{1}}}
\newcommand{\two}{{\sss{2}}}
\newcommand{\three}{{\sss{3}}}
\newcommand{\five}{{\sss{5}}}
\newcommand{\eleven}{\sss{11}}
\newcommand{\twelve}{{\sss{12}}}
\newcommand{\twentytwo}{{\sss{22}}}
\newcommand{\twentyone}{{\sss{21}}}

\newcommand{\calH}{\mathcal{H}}

\newcommand{\calF}{\mathcal{F}}

\newcommand{\calS}{\mathcal{S}}
\newcommand{\calL}{\mathcal{L}}

\newcommand{\calM}{\mathcal{M}}

\newcommand{\calHJF}{\jordan{\calH}}
\newcommand{\calP}{\mathcal{P}}

\newcommand{\ueq}{\mathrm{eq}}
\newcommand{\urad}{\mathrm{rad}}

\newcommand{\uini}{\mathrm{ini}}
\newcommand{\uquant}{\mathrm{q}}
\newcommand{\uend}{\mathrm{end}}
\newcommand{\usr}{\mathrm{sr}}
\newcommand{\ueff}{\mathrm{eff}}
\newcommand{\ucav}{\mathrm{cav}}
\newcommand{\upn}{\mathrm{pn}}
\newcommand{\ure}{\mathrm{re}}

\newcommand{\ud}{\mathrm{d}}
\newcommand{\ue}{\mathrm{e}}

\newcommand{\ub}{\mathrm{b}}
\newcommand{\uc}{\mathrm{c}}
\newcommand{\up}{\mathrm{p}}
\newcommand{\ur}{\mathrm{r}}
\newcommand{\uS}{\sss{\mathrm{S}}}
\newcommand{\uT}{\sss{\mathrm{T}}}

\newcommand{\efold}{n}
\newcommand{\bfold}{n - n_\uend}

\newcommand{\modP}{\varphi}
\newcommand{\modR}{\psi}
\newcommand{\mat}{\chi}
\newcommand{\thetaone}{\theta_\one}
\newcommand{\thetatwo}{\theta_\two}
\newcommand{\sone}{s_\one}
\newcommand{\stwo}{s_\two}
\newcommand{\Sone}{S_\one}
\newcommand{\Stwo}{S_\two}
\newcommand{\mukha}{Q}
\newcommand{\qmode}{\mu}
\newcommand{\qmodeS}{\qmode_\uS}
\newcommand{\qmodeT}{\qmode_\uT}
\newcommand{\source}{\calS}

\newcommand{\matnd}{{\bar{\mat}}}

\newcommand{\Vall}{V}
\newcommand{\Vgrav}{W}
\newcommand{\Vmat}{U}
\newcommand{\Zall}{Z}

\newcommand{\warp}{c_\mathrm{w}}
\newcommand{\kbulk}{k_\ub}
\newcommand{\baremass}{m_{\uc}}
\newcommand{\baremassnd}{\bar{m}_{\uc}}
\newcommand{\baremasseff}{\bar{m}_{\ueff}}
\newcommand{\cplP}{c_{\up}}
\newcommand{\cplR}{c_{\ur}}
\newcommand{\const}{C}
\newcommand{\constsr}{\const_\usr}
\newcommand{\constdec}{\const_\uquant}
\newcommand{\constconv}{\const_\uc}

\newcommand{\Afac}{A}
\newcommand{\alphafac}{\alpha}
\newcommand{\betafac}{\beta}
\newcommand{\kappaeff}{\kappa_\ueff}
\newcommand{\kappacav}{\kappa_\ucav}
\newcommand{\kappafive}{\kappa_\five}

\newcommand{\zone}{z_\one}
\newcommand{\ztwo}{z_\two}

\newcommand{\epsone}{\epsilon}
\newcommand{\epsoneJF}{\jordan{\epsone}}
\newcommand{\epstwo}{\epsilon_{\two}}

\newcommand{\aJF}{\jordan{a}}

\newcommand{\energy}{\rho}
\newcommand{\energyJF}{\jordan{\energy}}

\newcommand{\gammaPN}{\gamma_\upn}

\newcommand{\hubble}{H}
\newcommand{\hubblend}{\bar{\hubble}}
\newcommand{\hubbleJF}{\jordan{\hubble}}

\newcommand{\firstform}{\eta}
\newcommand{\unitv}{\iota}

\newcommand{\kwav}{k}
\newcommand{\kmpc}{\kwav_\zero}
\newcommand{\kpivot}{\kwav_*}

\newcommand{\adia}{\sigma}

\newcommand{\metric}{\ell}
\newcommand{\christoffel}{\Upsilon}
\newcommand{\field}{\calF}

\newcommand{\rotmat}{\calM}
\newcommand{\Vone}{\Vall_\one}
\newcommand{\Vtwo}{\Vall_\two}
\newcommand{\Veleven}{\Vall_\eleven}
\newcommand{\Vtwelve}{\Vall_\twelve}
\newcommand{\Vtwentyone}{\Vall_\twentyone}
\newcommand{\Vtwentytwo}{\Vall_\twentytwo}
\newcommand{\Vadia}{\Vall_\adia}
\newcommand{\Zeleven}{\Zall_\eleven}
\newcommand{\Vadiaadia}{\Vall_{\adia\adia}}
\newcommand{\Vadiaone}{\Vall_{\adia \one}}
\newcommand{\Voneadia}{\Vall_{\one \adia}}
\newcommand{\Vadiatwo}{\Vall_{\adia \two}}
\newcommand{\Vtwoadia}{\Vall_{\two \adia}}

\newcommand{\Phipert}{\Phi}
\newcommand{\Psipert}{\Psi}

\newcommand{\matndpert}{\delta \matnd}
\newcommand{\fieldpert}{\delta \field}
\newcommand{\adiapert}{\delta \adia}
\newcommand{\curvpert}{\zeta}

\newcommand{\sonepert}{\delta \sone}
\newcommand{\stwopert}{\delta \stwo}
\newcommand{\modPpert}{\delta \modP}
\newcommand{\modRpert}{\delta \modR}
\newcommand{\orthometric}{\perp}
\newcommand{\tens}{h}

\newcommand{\power}{\calP}
\newcommand{\obspert}{\nu}
\newcommand{\powerCurv}{\power_\curvpert}
\newcommand{\powerSone}{\power_{\Sone}}
\newcommand{\powerStwo}{\power_{\Stwo}}
\newcommand{\powerTens}{\power_\tens}

\newcommand{\OmegaJF}{\jordan{\Omega}}

\newcommand{\Mpc}{\mathrm{Mpc}}
\newcommand{\camb}{\textsc{camb}}
\newcommand{\cosmomc}{\textsc{cosmomc}}

\newcommand{\Omegab}{\Omega_\ub}
\newcommand{\Omegac}{\Omega_\uc}
\newcommand{\zre}{z_\ure}
\newcommand{\optdepth}{\tau}
\newcommand{\thetalewis}{\theta}
\newcommand{\like}{\calL}

\begin{document}

\title{Boundary Inflation and the WMAP Data}

\author{Christophe Ringeval}
\email{c.ringeval@imperial.ac.uk}
\affiliation{Blackett Laboratory, Imperial College, Prince
Consort Road, London SW7 2AZ, United Kingdom}

\author{Philippe Brax} \email{brax@spht.saclay.cea.fr}
\affiliation{Service de Physique Th\'eorique, Commissariat \`a
  l'\'Energie Atomique--Saclay, 91191 Gif-sur-Yvette Cedex, France}

\author{Carsten van de Bruck} \email{c.vandebruck@sheffield.ac.uk}
\affiliation{Department of Applied Mathematics, Sheffield University,
  Sheffield S3 7RH, United Kingdom}

\author{Anne-Christine Davis} \email{a.c.davis@damtp.cam.ac.uk}
\affiliation{Department of Applied Mathematics and Theoretical
  Physics, Center for Mathematical Sciences, University of Cambridge,
  Wilberforce Road, Cambridge CB3 OWA, United Kingdom}

\date{\today}

\begin{abstract}
  Inflation in a five-dimensional brane world model with two boundary
  branes is studied. We make use of the moduli space approximation
  whereby the low energy theory reduces to a four-dimensional
  biscalar-tensor gravity plus a minimally coupled scalar field. After
  a detailed analysis of the inflationary solutions, we derive the
  evolution equations of the linear perturbations separating the
  adiabatic mode from two entropy modes. We then examine the
  primordial scalar and tensor power spectra and show that their tilt
  depends on the scalar-tensor coupling constant. Finally, the induced
  CMB anisotropies are computed and we present a Monte Carlo Markov
  Chains exploration of the parameter space using the first year WMAP
  data. We find a marginalized probability bound for the associated
  Eddington parameter at the end of inflation $1-\gamma<2 \times
  10^{-3}$, at $95\%$ confidence level. This suggests that future CMB
  data could provide crucial information helping to distinguish
  scalar-tensor and standard inflationary scenarios.

\end{abstract}
\maketitle

\section{Introduction}

Although the standard particle physics and cosmological models in a
four-dimensional spacetime provide an accurate enough description of
particle physics experiments and cosmological
observations~\cite{PDBook}, considering unseen extra dimensions allows
one to make theoretical progress towards a Unified Theory of all
interactions~\cite{Nordstrom:1914, Kaluza:1921,
  Klein:1926, O'Raifeartaigh:1998pk, Polchinski:1998rq,
  Polchinski:1998rr}. The advent of M-theory and branes has led to the
``brane world'' concept in cosmology.  Motivated by the
Ho\v{r}ava--Witten model and its
compactifications~\cite{Horava:1996ma, Witten:1996mz}, brane world
models assume that we are living on a higher dimensional spacetime
boundary (brane).

In cosmology, brane worlds have been studied within the framework of
five-dimensional (5D) models and lead to only a few observable
predictions (see Refs~\cite{Langlois:2002bb, Maartens:2003tw,
  Brax:2004xh} for a review and references therein).  Models ranging
from classical membrane theories of General Relativity to topological
defect models have been studied in various compact or non-compact,
curved or flat, symmetric or non-symmetric, extra dimensional
spacetimes, as well as their associated deviations, or
incompatibilities, with respect to the standard
cosmology~\cite{Akama:1982jy, Rubakov:1983bb, Davis:2000jq,
  Battye:1995hv, Battye:2001yn, Binetruy:1999ut, Maartens:1999hf,
  Randall:1999vf, Randall:1999ee, Garriga:1999yh, Arkani-Hamed:1998rs,
  Antoniadis:1998ig, Ringeval:2003na, Dvali:2000hr,Antoniadis:2002tr,
  Cremades:2002dh,Kokorelis:2002qi, Kohlprath:2003pu,
  Kolanovic:2003am, Ringeval:2004ju, Cvetic:1996vr, Bonjour:1999kz,
  Antunes:2002hn, Peter:2003zg, Koyama:2004ap, Cartier:2004sq}.
Definite cosmological predictions remain however difficult to extract
in the high energy limit where the universe evolution is strongly
affected by the physical processes occurring in the extra dimensions.
On the other hand, at lower energy scales (typically smaller than the
brane tension) it appears possible to describe such systems by
effective four-dimensional actions.

In the framework of five-dimensional compactifications of
M-theory~\cite{Lukas:1999yn,Lukas:1998tt}, the moduli space
approximation describes, through a four-dimensional effective action,
a system of two branes of opposite tension embedded in a
five-dimensional warped spacetime~\cite{Brax:2000xk}. Besides the
fields living on the positive tension brane (assumed to be our
universe), the moduli associated with the position of the branes in
the fifth dimension act as two non-minimally coupled scalar fields
thereby leading to an effective biscalar-tensor theory of
gravity~\cite{Ashcroft:2002vj,Brax:2002nt, Kobayashi:2002pw,
  Brax:2003vf, Ashcroft:2003xx, Kanno:2003xy, Palma:2004fh,
  Davis:2005au}. Scalar-tensor theories have been intensively studied
as natural extensions of General Relativity and are strongly
constrained in the solar system~\cite{Damour:1992we,Will:2001mx}. On
the cosmological side, there are (weaker) constraints coming from the
Big Bang Nucleosynthesis (BBN) and recent works have been focused on
scalar-tensor effects in the Cosmic Microwave Background anisotropies
(CMB) and in weak gravitational lensing~\cite{Damour:1998ae,
  Perrotta:1999am, Chen:1999qh, Baccigalupi:2000je, Amendola:2000ub,
  Riazuelo:2001mg, Nagata:2003qn, Acquaviva:2004ti,
  Esposito-Farese:2004tu,Uzan:2004qr, Schimd:2004nq,Rhodes:2003ev}. In
these works, scalar-tensor theories have been discussed in the context
of the \emph{post-inflationary} cosmology by quantifying the
distortions the non-minimally scalar fields may induce compared to the
pure General Relativity predictions. In fact, even if the cosmological
constraints are currently weaker than the solar-system ones, they are
complementary in the sense that they probe the cosmic times. This is
relevant since it has been shown that scalar-tensor gravity models can
relax towards General Relativity during the cosmological
evolution~\cite{Damour:1993id}. As a result, such models may have no
observable impact in the post-inflationary eras, and in particular in
the solar system, while modifying significantly the dynamics of the
early universe.

On top of probing the post-inflationary eras, the precision reached by
the current CMB experiments allows one to constrain the shape of the
power spectra of the primordial cosmological perturbations. Recast in
the inflationary context, this is an opportunity to shed light on the
physics at earlier times and in particular the deviations
scalar-tensor theories may generate \emph{during} inflation.

In this paper, we are interested in the primordial inflationary eras
occurring in a boundary inflation model and their potential observable
effects in the CMB, given the first year Wilkinson Microwave
Anisotropies Probe (WMAP) data~\cite{Kogut:2003et,
  Verde:2003ey,Hinshaw:2003ex}. A minimal realistic three-fields model
is considered: two of them being the moduli associated with the
position of the branes in the extra dimension, and the other acts as a
standard inflaton field living on our brane and coupled to the matter
sector~\cite{Brax:2002nt,Kobayashi:2002pw, Brax:2003vf,
  Ashcroft:2003xx, Palma:2004et}. After a detailed analysis of the
several viable background evolutions, we compute and discuss the
primordial power spectra of the scalar and tensor perturbations
generated during inflation. We then derive the resulting CMB
anisotropies and perform a Monte Carlo Markov Chain exploration of the
parameter space in a regime where the deviations from General
Relativity predictions are dominated by the shape of the primordial
power spectra. By comparing the power spectra induced by the standard
chaotic inflaton field alone and the ones obtained by considering the
moduli, we find that the first year WMAP data can differentiate
between standard chaotic and boundary inflation. In fact, for a given
matter sector, the presence of non-minimally coupled scalar-fields
during inflation changes the tilt of the primordial power spectra due
to the running of the conformal factor. In our framework, we find the
presence of the moduli statistically disfavored by the data, either as
the WMAP data are insensitive to their effects when they are weakly
coupled, or as they lead to a strongly tilted power spectra. These
effects lead to an upper bound on the allowed values of the moduli
coupling constant during inflation, which can be recast in terms of
the more commonly used Eddington post-Newtonian
parameter~\cite{Will:2001mx}: $1-\gamma<2\times 10^{-3}$. On one hand,
this bound holds at the time of inflation and, according to the above
discussion, provides a very early constraint on scalar-tensor
theories. On the other hand, it relies on the running of the conformal
factor during inflation as well as the choice of the matter sector. In
particular, one may expect this bound to be modified, or evaded, with
more complicated potentials for the matter fields, or by freezing the
running of the conformal factor at the time where observable
cosmological perturbations were generated. Nevertheless, this result
suggests that the current and future CMB data can be used to probe
scalar-tensor theories as early as during the inflationary era.

In Sect.~\ref{sect:background}, the boundary inflation model is
introduced and the background inflationary evolutions solved both
analytically and numerically. Sect.~\ref{sect:linpert} is devoted to
the derivation of the linear perturbations around the background
evolution by means of a separation between the adiabatic and the two
entropy modes. The equations of motion obtained are solved numerically
and used to compute the resulting primordial power spectra at the end
of inflation. In Sect.~\ref{sect:cmb}, we use modified versions of the
public CMB codes $\camb$~\cite{Lewis:1999bs} and
$\cosmomc$~\cite{Lewis:2002ah}, linked with our inflationary code, to
probe the boundary model parameter space given the WMAP data. Finally,
the relevance and the possible improvements of our results are
discussed in the conclusion, and the possibility of using the CMB to
accurately probe scalar-tensor theories in the early universe is
raised.

\section{Background evolution}
\label{sect:background}

We will focus on the low energy effective theory of 5D brane world
models with a bulk scalar field and matter living on the positive
tension brane. In this context, the low energy degrees of freedom of
the model are the graviton leading to 4D gravity, the bulk scalar
field zero mode and the inter brane distance, \ie the radion. One can
also parametrise the two scalar modes as the two brane positions. In
the following we will denote by $\modP$ and $\modR$ the two moduli
which are related to the brane positions $\zone$ and $\ztwo$ by
\begin{equation}
\begin{aligned}
  \ue^\modP \cosh \modR & = \left(1-4 \kbulk \warp^2
    \zone\right )^{(2\warp^2+1)/4\warp^2},\\
  \ue^\modP \sinh \modR & = \left(1-4 \kbulk \warp^2
  \ztwo\right)^{(2\warp^2+1)/4\warp^2},
\end{aligned}
\end{equation}
where $\kbulk$ fixes the energy scale of the brane tensions and $\warp$
is a constant which determines how warped the extra dimension
is~\cite{Brax:2000xk,Brax:2002nt,Ashcroft:2003xx}. In particular for
$\warp =0$, we obtain the Randall--Sundrum case. The four-dimensional
(4D) gravitational coupling is related to the 5D gravitational
coupling via
\begin{equation}
  \kappa^2= \left(2\warp^2 +1 \right) \kbulk \kappafive^2.
\end{equation}
Using these relations one gets
\begin{equation}
\begin{aligned}
  \modP & = \dfrac{1}{2} \ln \left[ \left(1 - 4 \kbulk \warp^2
      \zone\right)^{(2\warp^2+1)/2\warp^2} \right. \\
  & \left. - \left(1 - 4 \kbulk \warp^2
      \ztwo\right)^{(2\warp^2+1)/2\warp^2} \right],
\end{aligned}
\end{equation}
and
\begin{equation}
  \tanh \modR = \left(\dfrac{1 - 4 \kbulk \warp^2 \ztwo}{1 - 4 \kbulk \warp^2
      \zone}\right)^{(2\warp^2+1)/4\warp^2}.
\end{equation}
Notice that $\modP \rightarrow -\infty$ and $\modR \rightarrow \infty$
when the two branes collide. On the contrary, $\modR \rightarrow 0$
and $\modP \rightarrow \infty$ correspond to the negative tension
brane being stuck at the bulk singularity $\ztwo= 1/4\warp^2 \kbulk$
and the positive tension brane receding away towards infinity. This
situation will happen during the inflationary evolution.

Denoting by $\mat$ the minimally coupled scalar field living on
our brane universe, the effective action in the Einstein frame,
with a metric of $(-,+,+,+)$ signature, reads
\begin{equation}
\label{eq:actionbrane}
\begin{aligned}
  S & = \dfrac{1}{2 \kappa^2 } \int \left[R - \cplP \left(\partial
      \modP \right)^2 - \cplR
    \left(\partial \modR \right)^2 - \Vgrav \right]  \sqrt{-g} \,\ud^4 x  \\
  & + \int \left[-\dfrac{1}{2} \Afac^2 \left(\partial \mat \right)^2 -
    \Afac^4\Vmat \right] \sqrt{-g} \, \ud^4 x ,
\end{aligned}
\end{equation}
where $\modP$ and $\modR$ stands for the moduli fields,
$\Afac(\modP,\modR)$ is the conformal factor, $\Vgrav(\modP,\modR)$
the moduli bulk potential and $\Vmat(\mat)$ the minimally coupled
scalar field potential on the brane. In the following, we will focus
on the minimal setup of Refs.~\cite{Brax:2002nt}, \ie
\begin{equation}
\label{eq:branesetup}
\begin{aligned}
  \Afac^2(\modP,\modR) & = \ue^{-(\cplP/3) \modP} \left(\cosh \modR
  \right)^{\cplR/3} ,\quad
  \Vgrav(\modP,\modR)  = 0,\\
  \cplP & = \dfrac{ 12 \warp^2}{1 + 2 \warp^2}, \quad \cplR =
  \dfrac{6}{1+2 \warp^2}.
\end{aligned}
\end{equation}
In terms of the 5D picture, $\warp$ parametrises the coupling of the
bulk scalar to the brane. Since the resulting four-dimensional theory
is of multiscalar-tensor kind, it is also convenient to introduce the
first conformal gradient
\begin{equation}
\label{eq:firstgrads}
\begin{aligned}
  \alphafac_\modP & \equiv \dfrac{\partial \ln \Afac}{\partial \modP}
  = -\dfrac{\cplP}{6}, \quad \alphafac_\modR \equiv \dfrac{\partial
    \ln \Afac}{\partial \modR} = \dfrac{\cplR}{6} \tanh \modR.
\end{aligned}
\end{equation}
These are the gravitational couplings of the two moduli. Their
present values are constrained by solar system experiments.
Moreover, we will assume a chaotic-like potential for $\mat$
\begin{equation}
\label{eq:potmat}
\Vmat(\chi) = \dfrac{1}{2} \baremass^2 \mat^2.
\end{equation}
In such a setup, the field $\mat$ is expected to be coupled to the
observed form of matter and it will be referred to as the
``matter field'' in the following. As in the standard chaotic model,
inflation can be driven by $\mat$, but also by the moduli as discussed
in the next section~\cite{Kofman:2004yc}.

\subsection{Basic equations}
\label{sect:bgequations}

\subsubsection{Sigma-model formalism}

For the sake of clarity it is more convenient to recast the action
(\ref{eq:actionbrane}) in terms of a non-linear
sigma-model~\cite{Damour:1992we,Damour:1993id,Koshelev:2005wk}
\begin{equation}
\label{eq:actionsigmodel}
\begin{aligned}
S & = \dfrac{1}{2 \kappa^2} \int \bigg[R - \metric_{ab} g^{\mu \nu}
  \partial_\mu \field^a \partial_\nu \field^b -
   2\Vall(\field^c) \bigg] \sqrt{-g} \, \ud^4 x,
\end{aligned}
\end{equation}
the field-manifold being defined through
\begin{equation}
\field^a =
\left(
\begin{array}{c}
\displaystyle
\matnd \\
\modP \\
\modR
\end{array}
\right), \quad
\metric_{ab} = \diag\left(\Afac^2,\cplP,\cplR \right),
\end{equation}
where the dimensionless scalar field $\matnd = \kappa \mat$ has been
introduced. {}From Eq.~(\ref{eq:branesetup}), the fields evolve in the
potential
\begin{equation}
\label{eq:potall}
\Vall = \kappa^2 A^4 \Vmat.
\end{equation}
In the following, it will be implicitly assumed that the Latin
indices $\{a,b,c,d\}$ refer to the field-manifold. Differentiating the
action (\ref{eq:actionsigmodel}) with respect to the metric leads to
the Einstein-Jordan equations
\begin{equation}
\label{eq:einstein}
\begin{aligned}
G_{\mu \nu} =\source_{\mu\nu},
\end{aligned}
\end{equation}
with the source terms
\begin{equation}
\label{eq:tmunu}
\source_{\mu\nu}=\metric_{ab} \source^{ab}_{\mu \nu} -
g_{\mu\nu} \Vall,
\end{equation}
where
\begin{equation}
  \source^{ab}_{\mu \nu} = \partial_{\mu}\field^a \partial_{\nu}
  \field^b - \dfrac{1}{2}
  g_{\mu\nu} \partial_\rho \field^a \partial^\rho \field^b.
\end{equation}
Similarly, the fields obey the Klein-Gordon-like equation
\begin{equation}
\label{eq:kleingordon}
\square \field^c + g^{\mu \nu} \christoffel^c_{ab} \partial_\mu \field^a
\partial_\nu \field^b = \Vall^c,
\end{equation}
where $\christoffel$ denotes the Christoffel symbol on the field-manifold
\begin{equation}
\christoffel^c_{ab} = \dfrac{1}{2} \metric^{cd} \left(\metric_{da,b} +
  \metric_{db,a} - \metric_{ab,d} \right),
\end{equation}
and $\Vall^c$ should be understood as the vector-like partial
derivative of the potential
\begin{equation}
\Vall^{c} = \metric^{cd} \Vall_d = \metric^{cd} \dfrac{\partial
  \Vall}{\partial \field^d}.
\end{equation}

\subsubsection{Equations of motion}

In a flat Friedman--\-Lema\^{i}tre--\-Robertson--\-Walker (FLRW)
Universe with metric
\begin{equation}
\label{eq:flrw}
\ud s^2 = g_{\mu \nu} \ud x^\mu \ud x^\nu = a^2(\eta) \left(-\ud \eta^2
+ \delta_{ij} \ud x^i \ud x^j \right),
\end{equation}
$\eta$ being the conformal time, the equations of motion
(\ref{eq:einstein}) and (\ref{eq:kleingordon}) simplify to
\begin{align}
\label{eq:einsteinTT}
3 \calH^2 & = \dfrac{1}{2} \metric_{ab} {\field}^{a}{}'
{\field}^{b}{}' +
a^2 \Vall , \\
\label{eq:einsteinIJ}
2 \calH' &+ \calH^2 = -\dfrac{1}{2} \metric_{ab} {\field^{a}}{}'
{\field^{b}}{}'+  a^2 \Vall, \\
\label{eq:modulicosmo}
{\field^c}{}'' &+ \christoffel^c_{ab} {\field^a}{}'
{\field^b}{}' +2 \calH {\field^c}{}' = - a^2 \Vall^c,
\end{align}
where a prime denotes differentiation with respect to the conformal
time and $\calH$ is the conformal Hubble parameter $\calH = a'/a$. In
the Einstein Frame, the ``new terms'' compared to minimally coupled
scalar multifield inflationary models are encoded in the sigma-model
metric $\metric_{ab}$ and the Christoffel symbols
$\christoffel^c_{ab}$.

Since we are interested in inflationary behaviour, it is more
convenient to work in terms of an ``efold'' time variable
\begin{equation}
\label{eq:efold}
\efold = \ln \left( \dfrac{a}{a_{\uini}} \right).
\end{equation}
Apart from being the relevant physical time quantity during an
inflationary era, when expressed in terms of $n$, the field equations
can be decoupled from the metric evolution~\cite{Collins:1971,
  Collins:1972, Shikin:1973, Damour:1992we, Damour:1993id} and
Eqs.~(\ref{eq:einsteinTT}) to (\ref{eq:modulicosmo}) separate into
\begin{align}
\label{eq:hubblesquare}
  &\hubble^2 = \dfrac{\Vall}{3 - \dfrac{1}{2}\dot{\adia}^2},\\
\label{eq:hubbledot}
  &\dfrac{\dot{\hubble}}{\hubble}  = -\dfrac{1}{2} \dot{\adia}^2, \\
\label{eq:fieldevol}
  &\dfrac{{\ddot{\field}^{c}} + \christoffel^c_{ab} \dot{\field}^{a}
    \dot{\field}^{b}}{3 - \dfrac{1}{2} \dot{\adia}^2}  +
  \dot{\field}^{c} = - \dfrac{\Vall^c}{\Vall}.
\end{align}
The dot denotes differentiation with respect to $n$, the physical
Hubble parameter is $\hubble = \calH/a$, and the field velocity (in
efold time) on the manifold is
\begin{equation}
\label{eq:adiadot}
\dot{\adia} = \sqrt{\metric_{ab} \dot{\field}^{a} \dot{\field}^{b}}.
\end{equation}
This is the derivative of the so-called adiabatic field. Indeed,
differentiating Eq.~(\ref{eq:hubbledot}) and using
Eq.~(\ref{eq:fieldevol}) yields the standard second order differential
equation for the adiabatic field $\adia$, \ie in conformal
time~\cite{Gordon:2000hv, DiMarco:2002eb}
\begin{equation}
\label{eq:adiaevol}
\adia'' + 2 \calH \adia' + a^2
\unitv^c \Vall_c = 0,
\end{equation}
where $\unitv^a$ are unit field-vectors along the classical trajectory
\begin{equation}
\label{eq:unitvector}
\unitv^a \equiv \dfrac{{\field^a}{}'}{\adia'}
= \dfrac{\dot{\field}^{a}} {\dot{\adia}}.
\end{equation}

As mentioned above, the fields evolution is decoupled from the Hubble
flow and Eq.~(\ref{eq:fieldevol}) mimics the evolution of a
``particle'' of variable inertia $1/(3 - \dot{\adia}^2/2)$ on a curved
manifold in presence of friction and an external force
$-\Vall^c/\Vall$~\cite{Damour:1992we,Damour:1993id}. {}From
Eqs.~(\ref{eq:branesetup}) and (\ref{eq:potmat}) one gets
\begin{equation}
\label{eq:forces}
\begin{aligned}
  \dfrac{\Vall^\modP}{\Vall} & = 4 \alphafac^\modP = -\dfrac{2}{3}, \\
  \dfrac{\Vall^\modR}{\Vall} & = 4 \alphafac^\modR = \dfrac{2}{3}
  \tanh
  \modR, \\
  \dfrac{\Vall^{\matnd}}{\Vall} & = \dfrac{1}{\Afac^2}
  \dfrac{\Vmat_\matnd}{\Vmat} = \dfrac{1}{\Afac^2} \dfrac{2}{\matnd},
\end{aligned}
\end{equation}
while the only non-vanishing sigma-model Christoffel symbols are
\begin{equation}
\label{eq:numchristoffel}
\begin{aligned}
  \christoffel^{\matnd}_{\matnd \modP} & = \alphafac_{\modP}, \quad
  \christoffel^{\modR}_{\matnd \matnd} = - \Afac^2 \alphafac^{\modR},
  \\ \christoffel^{\matnd}_{\matnd \modR} & = \alphafac_{\modR} , \quad
  \christoffel^{\modP}_{\matnd \matnd}  = - \Afac^2
  \alphafac^{\modP},
\end{aligned}
\end{equation}
where lowering and raising indices on the first conformal gradients is
performed with the field-manifold metric $\metric_{ab}$. The fact that
the effective potential for the fields is $\ln(\Vall)$ ensures that
the evolution of the fields in efold time is independent of any
multiplicative constant appearing in the definition of the matter
potential $\Vmat$, as $\baremass$ in the chaotic-like potential we are
interested in [see Eq.~(\ref{eq:potmat})]. Obviously, $\baremass$
still fixes the normalisation of the Hubble parameter in
Eq.~(\ref{eq:hubblesquare}) and is directly related to the amplitude
of the power spectrum of linear perturbations.

{}From Eqs.~(\ref{eq:fieldevol}) and (\ref{eq:forces}), we can draw the
qualitative evolution of the fields. Starting from reasonable initial
conditions, \ie $\dot{\adia}^2 < 6$ [see Eq.~(\ref{eq:hubblesquare})],
the friction terms will tend to produce a stationary regime in which
the fields relax toward the minimum of the potential $\ln(\Vall)$.
For instance, the dilaton field $\modR$ will relax toward vanishing
value as fast as
\begin{equation}
\dot{\modR} \simeq -\dfrac{2}{3} \quad \Rightarrow \quad \modR(\efold) \simeq
\modR_\uini - \dfrac{2}{3} \efold,
\end{equation}
and stays frozen at $\modR=0$ afterward, whereas $\modP$ grows as
\begin{equation}
\modP(\efold) \simeq \modP_\uini + \dfrac{2}{3} \efold.
\end{equation}
{}From Eq.~(\ref{eq:forces}), the matter field $\mat$ is expected to
slowly evolve from its initial value provided $\Afac^2 \matnd \gg 2$,
thereby allowing a period of slow-roll
inflation~\cite{Ashcroft:2002vj}. However, since the force term
involves also the conformal factor $\Afac^2$, the properties of the
inflationary era, as its duration and end, will be certainly dependent
on the evolution of the moduli fields $\modP$ and $\modR$.

In the next section, we will present some analytical approximations
and numerical solutions to the previous sketch.

\subsection{Inflationary solutions}
\label{sect:inflation}

\subsubsection{Scale factors acceleration}

Since we are dealing with a multiscalar-tensor theory, it is
convenient to clarify the definition of ``inflation'' in the
cosmological context. The frame where non-gravity measurements take
their usual form is the Jordan frame where matter is minimally coupled
to the metric. For a given conformal coordinate system, the metric
tensor $\jordan{g}$ in the Jordan frame is obtain from the metric
tensor $g$ in the Einstein frame by the conformal transformation
\begin{equation}
\label{eq:frame}
\jordan{g} = \Afac^2 g \quad \Rightarrow \quad \jordan{a} = \Afac a.
\end{equation}
As a result, the acceleration of the scale factor is not necessarily the
same in the Jordan and Einstein frame and may even take place in one
frame only~\cite{Esposito-Farese:2000ij}. To solve the flatness and
homogeneity problem, we are interested in inflationary era in the
Jordan frame, \ie when the first slow-roll parameter in that frame
\begin{equation}
\label{eq:epsoneJF}
\epsoneJF \equiv 1 - \dfrac{\calHJF'}{\calHJF^2} < 1.
\end{equation}
{}From Eq.~(\ref{eq:frame}), the conformal Hubble parameter in the
Jordan frame is also
\begin{equation}
\label{eq:hubbleJF}
\calHJF = \calH \left(1 + \alphafac_a \dot{\field}^{a} \right),
\end{equation}
where the dot denotes again the derivative with respect to $\efold$,
the total number of efold in the Einstein frame. {}From
Eq.~(\ref{eq:epsoneJF}), one gets
\begin{equation}
\label{eq:epsoneJFEF}
\epsoneJF = \dfrac{\epsone + \alphafac_a
    \dot{\field}^{a}}{1 + \alphafac_a
    \dot{\field}^{a}} - \dfrac{\alphafac_a \ddot{\field}^{a} +
  \betafac_{ab} \dot{\field}^{a} \dot{\field}^{b}}{\left(1 + \alphafac_a
    \dot{\field}^{a} \right)^2},
\end{equation}
where $\epsone$ is the first slow-roll parameter in the Einstein frame
and $\betafac_{ab}$ is the second conformal gradient defined as
\begin{equation}
\label{eq:secondgrads}
\betafac_{ab} \equiv \dfrac{\partial \alphafac_a}{\partial \field^b}.
\end{equation}
In our setup, from Eqs.~(\ref{eq:branesetup}) and
(\ref{eq:firstgrads}), its only non-vanishing value is
\begin{equation}
\label{eq:secondgradRR}
\betafac_{\modR\modR} = \dfrac{\cplR}{6} \dfrac{1}{\cosh^2 \modR},
\end{equation}
which rapidly vanishes for non-vanishing values of $\modR$. Moreover,
in a friction dominated regime $\ddot{\field}^{a} \simeq 0$, the
remaining term in Eq.~(\ref{eq:epsoneJFEF}) can be significant only at
the time when both $\betafac_{\modR\modR}$ and $\dot{\modR}$ are
non-vanishing, \ie when the $\modR$ field leaves the friction
dominated regime, where it was decreasing with constant velocity, to
be frozen on the attractor at $\modR=0$. As the result, we can expect
$\epsoneJF \simeq \epsone$ as long as the fields remain in the
friction dominated regimes.

\subsubsection{Minimal slow-roll approximations}

{}From the Einstein equation (\ref{eq:hubbledot}), the first slow-roll
parameter in the Einstein frame is simply related to the field
velocity by
\begin{equation}
\label{eq:epsoneVel}
\epsone = \dfrac{1}{2} \dot{\adia}^2 = \dfrac{1}{2} \Afac^2
\dot{\matnd}^2 + \dfrac{1}{2} \cplP \dot{\modP}^2 + \dfrac{1}{2} \cplR
\dot{\modR}^2.
\end{equation}
Putting everything together in Eq.~(\ref{eq:fieldevol}) yields
\begin{align}
 \label{eq:matndevol}
\ddot{\matnd} & = - \left(3 - \epsone - \dfrac{\cplP}{3} \dot{\modP} +
  \dfrac{\cplR}{3}  \dot{\modR} \tanh \modR \right) \dot{\matnd} -
\dfrac{2}{\Afac^2 \matnd} (3 - \epsone), \\
\label{eq:modPevol}
\ddot{\modP} & = - \left(3 - \epsone \right) \dot{\modP} -
\dfrac{1}{6} \Afac^2 \dot{\matnd}^2 + \dfrac{2}{3} (3-\epsone),\\
\label{eq:modRevol}
\ddot{\modR} & = -\left(3 -\epsone \right) \dot{\modR} +
\dfrac{\Afac^2 \dot{\matnd}^2}{6} \tanh \modR - \dfrac{2}{3} (3 -
\epsone) \tanh \modR.
\end{align}

The friction dominated evolution of the moduli fields discussed in the
previous section is recovered provided we assume the matter field to
be in slow-roll $\Afac \dot{\matnd} \ll 1$. In this limit,
Eqs.~(\ref{eq:modPevol}) and (\ref{eq:modRevol}) admit the solutions
\begin{equation}
\label{eq:modulispeed}
\dot{\modP} = \dfrac{2}{3}, \qquad \dot{\modR} = -\dfrac{2}{3},
\end{equation}
where it was assumed that $\modR > 1$. However, note that $\modR = 0$
is an exact solution of Eq.~(\ref{eq:modRevol}) and thus,
$\ddot{\modR}$ may be non-vanishing only during less than a few efolds
of transition between $\modR \simeq 1$ and $\modR = 0$. Let us point
out that the derivation of Eq.~(\ref{eq:modulispeed}) does not require
all the fields to be in slow-roll, and in particular $\epsone$ may not
be small~\cite{Linde:2001ae}. This is due to the simple form of the
conformal factor we are interested in, whereas in the generic case a
full slow-roll treatment may be required~\cite{Noh:2001ia,
  DiMarco:2005nq}. In the two regimes $\modR > 1$ and $\modR = 0$, the
conformal factor evolves as
\begin{equation}
\label{eq:confapprox}
\Afac^2 \simeq \Afac^2_\uini \ue^{-\const n},
\end{equation}
with $\const = 2 (\cplR + \cplP)/9$ for $\modR > 1$, and $\const =
2\cplP/9$ for $\modR = 0$. In both cases, since we are assuming $\Afac
\dot{\matnd} \ll 1$, we have $\epsone \simeq \const$ [see
Eq.~(\ref{eq:epsoneVel})]. The moduli are thus driving two expansion
eras with
\begin{equation}
\begin{aligned}
  \epsone & \simeq \dfrac{2}{9} \left(\cplR + \cplP \right) & (\modR >
  1), \qquad \epsone & \simeq \dfrac{2}{9} \cplP & (\modR = 0).
\end{aligned}
\end{equation}
In our setup, according to Eq.~(\ref{eq:branesetup}), the first era is
not inflation since $\epsone \simeq 4/3 > 1$, whereas the second one
is for values of the coupling constant $\cplP < 9/2$. Physically, it
means that starting from high values of $\modR$ only gives rise to a
period of non-inflationary expansion where the moduli $\modR$ relaxes
toward $\modR = 0$, and consequently, the effective Planck length
$\kappaeff = \Afac \kappa$ decreases exponentially. When $\modR$
reaches its attractor, an inflationary era driven by $\modP$ can
start, which also corresponds to a decreasing effective Planck length
[see Eq.~(\ref{eq:confapprox})].

In terms of the 5D picture, the fate of the moduli indicates that the
two branes move away from each other. The inflationary phase driven by
$\modP$ occurs once the right (negative tension) brane becomes close
enough to the bulk singularity. As $\chi$ is constant in this regime,
the energy density of the boundary inflaton is constant leading to a
constant detuning of the left brane tension. As already studied in
\cite{Brax:2000xk}, this leads to an accelerated phase with an
exponential potential and power law inflation. More details about this
power law inflation phase can be found in Ref.~\cite{Brax:2003vf}.

This picture remains valid provided $\Afac \dot{\matnd} \ll 1$ and
from Eqs.~(\ref{eq:matndevol}) and (\ref{eq:modulispeed}) we can check
the consistency of this approximation. {}From Eq.~(\ref{eq:confapprox}),
if $\matnd$ starts from large enough values such that $\Afac_\uini
\matnd_\uini \gg \sqrt{2/\epsone}$ then an approximate solution of
Eq.~(\ref{eq:matndevol}) is
\begin{equation}
\label{eq:matndevolstart}
\matnd - \matnd_\uini \simeq  - \dfrac{2}{ \epsone \Afac^2_\uini
  \matnd_\uini}
\ue^{\epsone n}.
\end{equation}
In this case, we have indeed
\begin{equation}
  \Afac \dot{\matnd} = \dfrac{2}{\Afac_\uini \matnd_\uini}
  \ue^{\epsone n /2} \ll 1,
\end{equation}
and the matter field $\matnd$ remains frozen at its initial value
until $\epsone \Afac^2 \matnd_\uini^2 \simeq 1$, \ie during
approximately
\begin{equation}
\label{eq:nefoldmodP}
\Delta n \simeq \dfrac{1}{\epsone} \ln\left(\Afac^2 \matnd_\uini^2
  \epsone \right)
\end{equation}
efolds after which Eq.~(\ref{eq:matndevolstart}) is no longer valid.
Then the field $\matnd$ starts to evolve and $\Afac \dot{\matnd}$
increases until $\epsone = 1$ for which inflation ends. Nevertheless,
the standard slow-roll approximation can be used to derive the
behavior of the matter field during this last phase. For reasonable
small value of $\cplP$, $\epsone$ ends up being dominated by the
matter field evolution [see Eq.~(\ref{eq:epsoneVel})] whereas the
behavior of the moduli remains almost the same. Indeed, as can be seen
in Eq.~(\ref{eq:modPevol}), the value of $\dot{\modP}$ will be
affected up to $10\%$ only by the matter field when $\Afac
\dot{\matnd} \simeq \sqrt{2}$, which is also the end of inflation [see
Eq.~(\ref{eq:epsoneVel})]. As a result, Eq.~(\ref{eq:confapprox}) is
still a good approximation for the conformal factor, and using the
slow-roll approximation $\ddot{\matnd} \ll \dot{\matnd}$ in
Eq.~(\ref{eq:matndevol}) leads to
\begin{equation}
\label{eq:matndevolsr}
\matnd^2 \simeq \matnd_\usr^2 - \dfrac{4}{\Afac^2_\uini}
\left(\dfrac{\ue^{\const \efold}}{\const} - \dfrac{\ue^{\const
      \efold_\usr}}{\const} \right),
\end{equation}
where the ``sr'' indices label the efold at which $\epsone \simeq
\Afac^2 \dot{\matnd}^2/2$. The end of inflation occurs for $\Afac^2
\dot{\matnd}^2 \simeq 2$ which gives the number of efolds in this last
stage
\begin{equation}
\label{eq:nefoldmatnd}
\Delta \efold_\usr \simeq \dfrac{1}{\const} \ln \left( \dfrac{1+
    \dfrac{\const}{4} \Afac_\uini^2 \matnd_\usr^2}{1 +
    \dfrac{\const}{2}} \right).
\end{equation}
Note that in the limit $\const \rightarrow 0$, we recover the usual
one-field slow-roll expression for a chaotic potential~\cite{Martin:2004um}
\begin{equation}
  \label{eq:nefoldchaotic}
  \Delta \efold_\usr \simeq \dfrac{1}{4} \kappaeff^2 \mat_\usr^2 -
  \dfrac{1}{2},
\end{equation}
with $\kappaeff = \Afac_\uini \kappa$.

\subsubsection{Numerical approach}

\begin{figure}
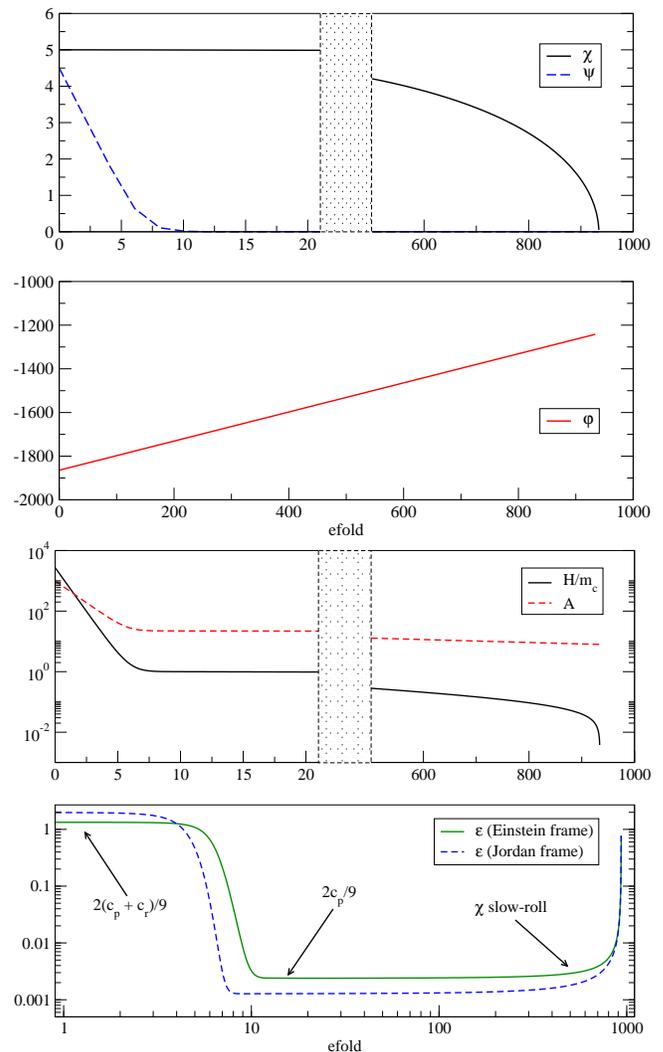

\begin{center}
\includegraphics[width=8.5cm]{fields_3p.eps}
\includegraphics[width=8.5cm]{geom_3p.eps}
\caption{Typical behaviors of the matter field $\matnd$ and the two
  moduli $\modP$ and $\modR$ plotted with respect to the total number
  of efold. The Hubble parameter in unit of the matter field bare mass
  $\baremass$, and the conformal factor, as well as the first
  slow-roll parameter in Einstein and Jordan frame are shown. As seen
  from the $\epsone$ behavior, this is a three stages expansion.  The
  model parameters are $\cplP=0.01$, $\cplR = 6 - \cplP$ [see
  Eq.~(\ref{eq:branesetup})], while the initial conditions have been
  chosen such that $\modR_\uini=4.5$, $\Afac_\uini = 1000$ and
  $\matnd_\uini = 5$. Moreover, the field velocities have been chosen
  to start immediately in the friction dominated regime, \ie with
  $\dot{\field}^{a}=-\Vall^a/\Vall$ (see also
  Fig.~\ref{fig:velocities}).}
\label{fig:3phases}
\end{center}
\end{figure}

A numerical integration of the equations of motion
(\ref{eq:hubblesquare}) to (\ref{eq:fieldevol}) has been performed and
illustrates the analytical behavior described in the previous section.
In Fig.~\ref{fig:3phases}, the three expansion eras can be
distinguished by means of the $\epsone$ values. For the chosen model
parameters, during the first ten efolds, the field $\modR$ relaxes
towards zero and its evolution dominates the expansion. As previously
pointed out, this is not an inflationary era in our model. During this
phase, the conformal factor, and thus the Hubble parameter, decrease
exponentially [see Eqs.~(\ref{eq:potall}) and
(\ref{eq:hubblesquare})].

Once $\modR$ reaches its  minimum, the dynamics are  dominated by
the second moduli $\modP$ and $\epsone \simeq 2 \cplP/9$. Note
that during the transition between these two regimes, the
slow-roll parameter $\epsone$ and $\epsoneJF$, in the Einstein and
Jordan frames respectively, do not match, as expected from
Eq.~(\ref{eq:epsoneJFEF}). As can be seen on
Fig.~\ref{fig:3phases}, inflation (in the Jordan frame) starts a few
efolds before $\epsone < 1$ in the Einstein frame. During the
$\modP$ dominated inflationary era, the matter field $\matnd$
leaves its initial value and starts to slow-roll under the
external force $-\Vall^{\matnd}/\Vall$ [see Eq.~(\ref{eq:forces})]
leading to the observed variation of $\epsone$ after $\efold\simeq
100$. The last stage occurs when the inflation is mainly driven by
$\matnd$ and we recover a one-field like behavior until $\epsoneJF
= 1$ for which inflation stops. Note that, as expected from the
analytical study, even in this last regime, the evolution of the
moduli $\modP$ and $\modR$ is not significantly affected by the
rolling of the matter field. As a result, their second
derivatives remain small and $\epsoneJF \simeq \epsone$ at the end
of inflation (see Fig.~\ref{fig:3phases}).
\begin{figure}
\begin{center}
\includegraphics[width=8.5cm]{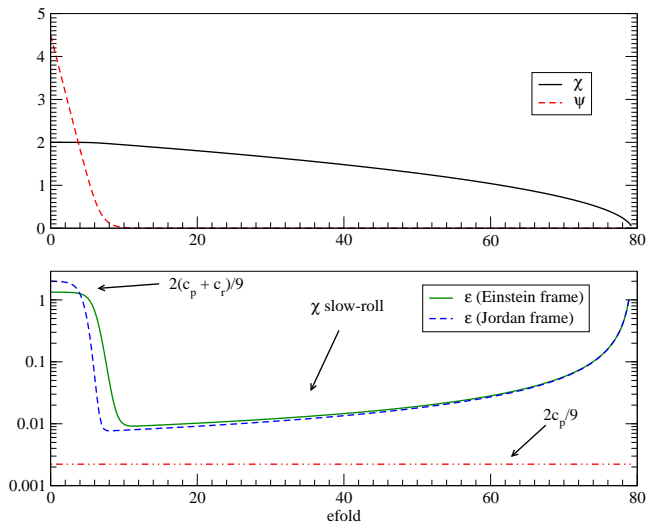}
\caption{Evolution of the fields for a two stages expansion. The
  matter field $\matnd$ is destabilized as soon as $\modR$ reaches its
  attractor $\modR=0$. As the evolution of $\epsone$ shows, there is
  not time for a pure $\modP$ dominated inflationary era and we start
  immediately in a $\matnd$ slow-roll dominated inflationary era until
  $\epsone=1$. These solutions have been obtained for the same model
  parameters as in Fig.~\ref{fig:3phases}, \ie $\cplP=0.01$, and for
  the initial conditions $\modR_\uini=4.5$, $\Afac_\uini=400$ and
  $\matnd_\uini=2$.}
\label{fig:2phases}
\end{center}
\end{figure}
\begin{figure}
\begin{center}
\includegraphics[width=8.5cm]{velocities.eps}
\caption{Relaxation toward the friction dominated regime. The dash and
  thick curves represent the evolution of the field velocities (in
  efold time) and the slow-roll parameter for initial conditions such
  that $\modR_\uini=0$, $\Afac_\uini=1$, $\matnd_\uini=16$,
  $\cplP=0.01$, with arbitrary chosen initial field velocities:
  $\dot{\matnd}_\uini=1$, $\dot{\modP}_\uini=-0.5$ and
  $\dot{\modR}_\uini=0.6$. The dotted curves are obtained for the same
  model parameters and initial field values, but with initial
  velocities on the attractor $\dot{\field}^{a} = -\Vall^a/\Vall$. As
  expected, it takes less than a few efolds for the velocities to relax
  toward their friction dominated values. Note the difference between
  $\epsoneJF$ and $\epsone$ during the relaxation time.}
\label{fig:velocities}
\end{center}
\end{figure}
\begin{figure}
\begin{center}
\includegraphics[width=8.5cm]{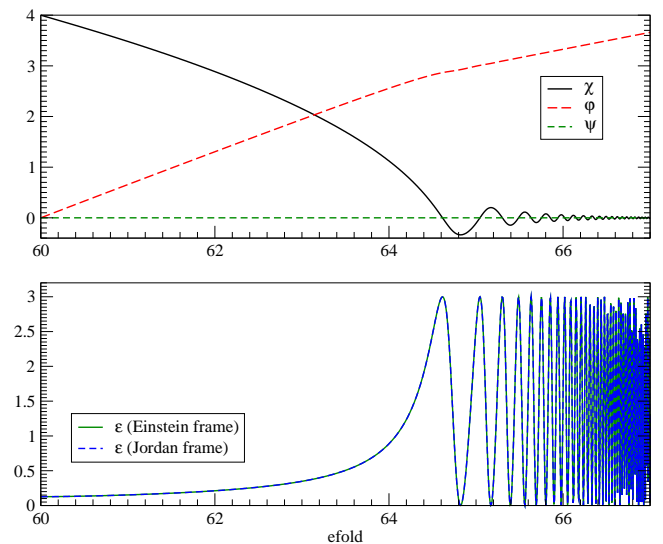}
\caption{The end of inflation and the beginning of the reheating phase
  where the matter field oscillates. Note the new slope
  $\dot{\modP}\simeq 1/3$ coming from the average value of the
  slow-roll parameter during that stage [see Eq.~(\ref{eq:modPevol})].
  Since $\modP$ and $\modR$ are not dramatically affected by the
  oscillations, one still have $\epsoneJF \simeq \epsone$.}
\label{fig:endinf}
\end{center}
\end{figure}
The above settings can nevertheless be affected according to the
parameter values and the initial conditions. Indeed, the matter
field $\matnd$ can leave its initial value as soon as, or even
before, the end of the $\modR$ domination expansion. In that case,
as can be seen in Fig.~\ref{fig:2phases}, there is no time for a
pure $\modP$ dominated inflation and after a intermediate mixed
phase, we jump directly to the last $\matnd$ dominated stage.

Concerning the initial conditions, it is important to mention that we
have fixed the initial derivatives of the fields in such a way that
their evolution start in the friction dominated regime, \ie with an
initial boost $\dot{\field}^{a} = - \Vall^a/\Vall$. In
Fig.~\ref{fig:velocities}, we have plotted the behavior of the fields
obtained for an arbitrary choice of reasonable initial velocities,
\ie verifying $\dot{\adia}^2 < 6$. As expected, it takes less than few
efolds for the fields to relax toward their attractor behavior. As
previously discussed, during the relaxation time, the slow-roll
parameters in Einstein and Jordan frame are significantly different.

In order to derive the primordial perturbations produced in this model
and the resulting cosmic microwave background anisotropies (CMB), it
is necessary to assume a cosmological scenario for the background.
Motivated by the brane world picture where the moduli $\modP$ and
$\modR$ are related to the position of the branes in the
five-dimensional spacetime, it is reasonable to consider that these
two fields do not decay and remain present in the late-time cosmology.
As a result, only $\matnd$ can decay into radiation and dark matter at
the end of inflation. In Fig.~\ref{fig:endinf}, we have plotted the
fields evolution during a few efolds after the end of inflation, the
matter field $\matnd$ ends up oscillating thereby starting a period of
reheating from which the radiation dominated era can
begin~\cite{Turner:1983he,Shtanov:1994ce,Kofman:1997yn}. Note that
even if inflation can be driven by $\modP$ (see
Sect.~\ref{sect:inflation}), the presence of $\matnd$ is still
required to end it. The observational consequences coming from these
assumptions will be more detailed in Sect.~\ref{sect:cmb}.

In the next section, the linear perturbations arising during the
inflationary eras are discussed, as well as their primordial power
spectra.

\section{Linear perturbations}
\label{sect:linpert}

\subsection{Gravitational and matter sector}

In the longitudinal gauge, the scalar perturbations (with respect to
the rotations of the three-dimensional space) of the FLRW metric can
be expressed as
\begin{equation}
\label{eq:pertscalmetric}
\ud s^2 = a^2 \left[ - \left(1 + 2 \Phipert \right) \ud \eta^2 +
  \left(1 - 2 \Psipert \right) \gamma_{ij} \ud x^i \ud x^j \right],
\end{equation}
where the indices $i$ and $j$ refer to the spatial coordinates only,
and $\Phipert$ and $\Psipert$ are the Bardeen potentials. In the
sigma-model formalism, denoting by $\fieldpert$ the scalar
perturbations of the fields, the Einstein-Jordan equations perturbed
at first order are
\begin{align}
\label{eq:energypert}
3 \calH \Psipert' & + \left(\calH'+2\calH^2 \right)\Psipert - \Delta
\Psi = - \dfrac{1}{2} \metric_{ab} {\field^a}{}' {\fieldpert^b}{}'
\nonumber \\ & - \dfrac{1}{2} \left(\dfrac{1}{2}\metric_{ab,c} {\field^a}{}'
  {\field^b}{}' + a^2 \Vall_c \right) \fieldpert^c,  \\
\label{eq:momentumpert}
\Psipert'& + \calH \Psipert = \dfrac{1}{2} \metric_{ab} {\field^a}{}'
\fieldpert^b, \\
\label{eq:diagonalpert}
\Psipert'' &+ 3\calH \Psipert' + \left(\calH' + 2\calH^2 \right)
\Psipert = \dfrac{1}{2} \metric_{ab} {\field^a}{}' {\fieldpert^b}{}'
\nonumber \\ & + \dfrac{1}{2} \left(\dfrac{1}{2} \metric_{ab,c} {\field^a}{}'
  {\field^b}{}' - a^2 \Vall_c \right) \fieldpert^c ,
\end{align}
where the prime stands for the derivative with respect to the
conformal time and use has been made of $\Phipert=\Psipert$ (perturbed
Einstein equation for $i\ne j$). The Laplacian reduces to
\begin{equation}
\Delta \equiv \delta^{ij} \partial_i \partial_j,
\end{equation}
for flat spatial hypersurfaces. The perturbed equations of motion for
the fields stem from Eq.~(\ref{eq:kleingordon}) and read
\begin{align}
  \label{eq:fieldpert} {\fieldpert^c}{}'' & + 2 \christoffel^c_{ab}
  {\field^a}{}' {\fieldpert^b}{}' + 2 \calH {\fieldpert^c}{}' \nonumber
  \\ & + \Bigg(\christoffel^c_{ab,d} {\field^a}{}'{\field^b}{}' +
  a^2\Vall^c_d -
  \metric^{ca}\metric_{ab,d} \, a^2 \Vall^b \Bigg) \fieldpert^d \nonumber \\
  & - \Delta \fieldpert^c = 4 \Psipert' {\field^c}{}' - 2\Psipert a^2 \Vall^c.
\end{align}

The fact that there is more than one scalar field involved leads to
the existence of entropy modes that can source the adiabatic mode. In
the next section, we use the formalism developed in
Refs.~\cite{Gordon:2000hv,GrootNibbelink:2001qt, Hwang:2001fb,
  DiMarco:2002eb} to derive the evolution equations of the adiabatic
and the two entropy modes arising in our model.

\subsection{Adiabatic and entropy perturbations}

\subsubsection{Generic decomposition}

{}From Eqs.~(\ref{eq:energypert}) to (\ref{eq:diagonalpert}), up to the
background equations, the evolution of the Bardeen potential
simplifies to
\begin{equation}
\label{eq:bardeenevol}
\begin{aligned}
\Psipert'' + 6 \calH \Psipert' + \left(2 \calH' + 4 \calH^2 \right)
\Psipert - \Delta \Psipert = - a^2 \Vall_c \fieldpert^c.
\end{aligned}
\end{equation}
In terms of the comoving curvature perturbation~\cite{Lukash:1980iv,
  Lyth:1984gv, Kodama:1985bj, Mukhanov:1990me, Durrer:1993db,
  Martin:1997zd}
\begin{equation}
\label{eq:curvpertdef}
\curvpert \equiv \Psipert - \dfrac{\calH}{\calH' - \calH^2} \left(
\Psipert' + \calH \Phipert \right),
\end{equation}
Eq.~(\ref{eq:momentumpert}) yields
\begin{equation}
\label{eq:curvpertadiab}
\curvpert= \Psipert + \calH \dfrac{\adiapert}{\adia'},
\end{equation}
where the adiabatic perturbation $\adiapert$ is the perturbed version
of Eq.~(\ref{eq:adiadot}) as well as the resulting perturbation of all
fields projected onto the classical trajectory [see
Eq.~(\ref{eq:unitvector})]:
\begin{equation}
\adiapert = \dfrac{\metric_{ab} {\field^a}{}'
  \fieldpert^b}{\adia'} = \unitv_a \fieldpert^a.
\end{equation}
The dynamical equation (\ref{eq:bardeenevol}) also reads
\begin{equation}
\curvpert' = \dfrac{2\calH}{\adia'^2} \Delta \Psipert -
\dfrac{2\calH}{\adia'^2} \left(a^2 \Vall_a \fieldpert^a -
  a^2\dfrac{\Vall_c {\field^c}{}'}{\adia'} \dfrac{\metric_{ab}
    {\field^a}{}' \fieldpert^b}{\adia'} \right),
\end{equation}
which can be recast into
\begin{equation}
\label{eq:curvpertevol}
\curvpert' = \dfrac{2\calH}{\adia'^2} \Delta \Psipert -
\dfrac{2\calH}{\adia'^2}  \orthometric^c_d a^2 \Vall_c \fieldpert^d.
\end{equation}
The orthogonal projector is defined by
\begin{equation}
\orthometric_{ab} = \metric_{ab} - \firstform_{ab},
\end{equation}
where $\firstform_{ab}\equiv \unitv_a \unitv_b$ is the first
fundamental form of the one-dimensional manifold defined by the
classical trajectory~\cite{Carter:1997pb}. We recover that the
comoving curvature perturbation on super-Hubble scales ($\Delta
\Psipert \simeq 0$) is only sourced by the entropy perturbations
defined as the projections of all field perturbations on the
field-manifold subspace orthogonal to the classical
trajectory~\cite{Martin:1997zd}.

\subsubsection{Spherical basis}

For the studied boundary inflation model, the sigma-model manifold is
three-dimensional which implies the existence of two entropy modes
living in the two-dimensional subspace orthogonal to the classical
trajectory. The decomposition performed in
Refs.~\cite{Gordon:2000hv,DiMarco:2002eb} is straightforwardly
generalized by choosing a local spherical basis at each point of the
fields trajectory. This can be performed by introducing the angular
fields $\thetaone$ and $\thetatwo$ defined by
\begin{equation}
\label{eq:thetaonedef}
\begin{aligned}
\cos \thetaone &\equiv \dfrac{\Afac \matnd'}{\adia'}, \qquad \sin
    \thetaone \equiv \dfrac{1}{\adia'} \sqrt{\cplP
    \modP'^2 + \cplR \modR'^2},
\end{aligned}
\end{equation}
and
\begin{equation}
\label{eq:thetatwodef}
\begin{aligned}
\cos
    \thetatwo &\equiv \dfrac{\sqrt{\cplP} \modP'}{\sqrt{\cplP
    \modP'^2 + \cplR \modR'^2}}, \qquad
\sin \thetatwo \equiv \dfrac{\sqrt{\cplR} \modR'}{\sqrt{\cplP
    \modP'^2 + \cplR \modR'^2}},
\end{aligned}
\end{equation}
provided $\modP'$ and $\modR'$ do not vanish at same times. They
define an instantaneous rotation matrix on the field-manifold
\begin{equation}
\label{eq:rotmat}
\rotmat=
\left(
\begin{array}{ccc}
\cos \thetaone & \sin \thetaone \cos \thetatwo & \sin \thetaone \sin
\thetatwo \\ -\sin \thetaone & \cos \thetaone \cos \thetatwo & \cos
\thetaone \sin \thetatwo \\ 0 & -\sin \thetatwo & \cos \thetatwo
\end{array}
\right),
\end{equation}
which transforms the original fields into the adiabatic and entropy
modes

\begin{equation}
\label{eq:oldperttonewpert}
\left(
\begin{array}{c}
\adiapert \\
\sonepert \\
\stwopert
\end{array}
\right) \equiv \rotmat
\left(
\begin{array}{c}
\Afac \matndpert \\
\cplP^{1/2} \modPpert \\
\cplR^{1/2} \modRpert
\end{array}
\right).
\end{equation}
The evolution of the angular fields is readily obtained by
differentiating Eqs.~(\ref{eq:thetaonedef}) and (\ref{eq:thetatwodef})
and using the background equations (\ref{eq:einsteinTT}) to
(\ref{eq:modulicosmo})
\begin{equation}
\label{eq:thetaoneevol}
\begin{aligned}
\thetaone' & = -\dfrac{a^2 \Vone}{\adia'} + \adia'
\cos \thetaone \left(\dfrac{\alphafac_\modP}{\sqrt{\cplP}}
 \cos \thetatwo +
\dfrac{\alphafac_\modR}{\sqrt{\cplR}}  \sin
\thetatwo \right ),\\
\end{aligned}
\end{equation}
\begin{equation}
\label{eq:thetatwoevol}
\begin{aligned}
\sin \thetaone \thetatwo' & = -\dfrac{a^2 \Vtwo}{\adia'}
 - \adia' \cos^2 \thetaone \left(\dfrac{\alphafac_\modP}{\sqrt{\cplP}}
 \sin \thetatwo -
\dfrac{\alphafac_\modR}{\sqrt{\cplR}} \cos \thetatwo \right ),
\end{aligned}
\end{equation}
for $\thetaone$ and $\thetatwo$, respectively. The rotated potential
derivatives have been defined through
\begin{equation}
\label{eq:oldpottonewpot}
\left(
\begin{array}{c}
\Vadia \\
\Vone \\
\Vtwo
\end{array}
\right) \equiv \rotmat
\left(
\begin{array}{c}
\Vall_\matnd/\Afac \\
\Vall_\modP/\cplP^{1/2}  \\
\Vall_\modR/\cplR^{1/2}
\end{array}
\right).
\end{equation}
Note that since we are dealing with non-minimally coupled scalar
fields, they are not the partial derivatives of the potential $\Vall$
with respect to the rotated fields~\cite{DiMarco:2002eb}, however
$\Vadia = \unitv^a \Vall_a$ remains the effective potential which
sources the adiabatic field [see Eq.~(\ref{eq:adiaevol})]. In the
spherical basis, we recover explicitly that the entropy modes only
source the comoving curvature perturbation since in
Eq.~(\ref{eq:curvpertevol}) one has $\orthometric^c_d \Vall_c \fieldpert^d =
\Vone \sonepert + \Vtwo \stwopert$. In order to compute the primordial
power spectra and the cross-correlation for the different modes, it is
convenient to recast the equations of motion for the original fields
in terms of the rotated modes
only~\cite{Langlois:1999dw,Gordon:2000hv,DiMarco:2002eb}.
\vspace{0.7cm}
\subsubsection{Equations of motion}

The closed system of dynamical equations for the entropy and adiabatic
modes can be obtained by expressing the second order derivative of
each mode with respect to the conformal time in terms of the others by
means of Eqs.~(\ref{eq:oldperttonewpert}) and
(\ref{eq:oldpottonewpot}), using Eqs.~(\ref{eq:einsteinTT}) to
(\ref{eq:modulicosmo}) as well as Eqs.~(\ref{eq:thetaoneevol}) and
(\ref{eq:thetatwoevol}) and their derivatives.  Moreover, the
adiabatic field $\adiapert$ can be expressed in terms of the comoving
curvature perturbations by means of Eq.~(\ref{eq:curvpertadiab}),
which is a preferred observable for deriving the subsequent Cosmic
Microwave Background (CMB) anisotropies (see Sect.~\ref{sect:cmb}).
After straightforward but tremendous calculations one gets for the first
entropy mode $\sonepert$
\begin{widetext}
\vspace{-0.7cm}
\begin{equation}
\label{eq:sonepertevol}
\begin{aligned}
  \sonepert'' & + 2 \calH \sonepert' + \dfrac{2\adia'}{\tan \thetaone}
  \left(\dfrac{a^2 \Vtwo}{\adia'^2} +
    \dfrac{\alphafac_\modP}{\sqrt{\cplP}} \sin \thetatwo - \dfrac{
      \alphafac_\modR}{\sqrt{\cplR}} \cos \thetatwo \right) \stwopert'
  + \Bigg[-\Delta + a^2 \Zeleven + \dfrac{11 + 4 \cos 2\thetaone +
    \cos 4\thetaone}{8 \sin^2 \thetaone} a^2 \Vtwo \\ & \times
  \left(-\dfrac{\alphafac_\modP}{\sqrt{\cplP}} \sin \thetatwo +
    \dfrac{ \alphafac_\modR}{\sqrt{\cplR}} \cos \thetatwo \right) +
  \sin \thetaone a^2 \Vadia
  \left(\dfrac{\alphafac_\modP}{\sqrt{\cplP}} \cos \thetatwo +
    \dfrac{\alphafac_\modR}{\sqrt{\cplR}}\sin \thetatwo \right) -
  \dfrac{a^4 \Vone^2}{\adia'^2} - \dfrac{a^4 \Vtwo^2}{\adia'^2 \tan^2
    \thetaone} \\ & - \adia'^2 \left(
    \dfrac{\betafac_{\modP\modP}}{\cplP} \cos^2 \thetatwo +
    \dfrac{\betafac_{\modR\modR}}{\cplR} \sin^2 \thetatwo +
    \dfrac{\betafac_{\modP\modR}}{\sqrt{\cplP \cplR}} \sin 2 \thetatwo
  \right) - \dfrac{\adia'^2}{\tan^2 \thetaone} \left( -
    \dfrac{\alphafac_{\modP}}{\sqrt{\cplP}} \sin \thetatwo +
    \dfrac{\alphafac_\modR}{\sqrt{\cplR}} \cos \thetatwo \right)^2 \\
  & - \adia'^2 \left( \dfrac{\alphafac_\modP}{\sqrt{\cplP}} \cos
    \thetatwo + \dfrac{ \alphafac_\modR}{\sqrt{\cplR}} \sin \thetatwo
  \right)^2 \Bigg] \sonepert + \Bigg\{2 a^2 \Vtwelve + \left(\dfrac{15
      - 8 \cos 2\thetaone + \cos 4 \thetaone}{4 \sin^2 \thetaone} a^2
    \Vone \right. \\ & \left. + \dfrac{-9 \cos \thetaone + \cos
      3\thetaone}{2 \sin \thetaone} a^2 \Vadia \right) \left(
    \dfrac{\alphafac_\modP}{\sqrt{\cplP}} \sin \thetatwo -
    \dfrac{\alphafac_\modR}{\sqrt{\cplR}} \cos \thetatwo \right) -
  \dfrac{2a^2 \Vtwo}{\sin \thetaone \tan \thetaone} \left( \dfrac{
      \alphafac_\modP}{\sqrt{\cplP}} \cos \thetatwo + \dfrac{
      \alphafac_\modR}{\sqrt{\cplR}} \sin \thetatwo \right) \\ & +
  \dfrac{2 a^2 \Vtwo}{\adia'^2} \left( \dfrac{ a^2 \Vone}{\tan^2
      \thetaone} + \dfrac{3 \calH \adia' + a^2 \Vadia}{\tan \thetaone}
  \right) + \dfrac{\sigma'^2 \cos \thetaone}{\tan^2 \thetaone} \left[
    \left(-\dfrac{\alphafac_\modP^2}{\cplP} +
      \dfrac{\alphafac_\modR^2}{\cplR} \right) \sin 2\thetatwo +
    \dfrac{2 \alphafac_\modP \alphafac_\modR }{\sqrt{\cplP\cplR}} \cos
    2\thetatwo \right] \\ & + \adia'^2 \cos \thetaone \left[
    \left(\dfrac{\betafac_{\modP \modP}}{\cplP} -
      \dfrac{\betafac_{\modR \modR}}{\cplR} \right) \sin 2 \thetatwo -
    \dfrac{2 \betafac_{\modP \modR}}{\sqrt{\cplP\cplR}} \cos
    2\thetatwo \right] \Bigg\} \stwopert = \dfrac{2 a^2 \Vone}{\calH}
  \curvpert',
\end{aligned}
\end{equation}
where use has been made of
\begin{equation}
\adiapert'= \dfrac{2 \Delta \Psipert}{\adia'} +
\adia' \Psipert + \left(\dfrac{\adia''}{\adia} - \calH
\right) \adiapert - \dfrac{2 a^2}{\adia'} \left(\Vone \sonepert
+ \Vtwo \stwopert \right),
\end{equation}
and, from Eq.~(\ref{eq:curvpertevol}),
\begin{equation}
\dfrac{\Delta \Psipert}{\adia'^2} = \dfrac{\curvpert'}{2
  \calH} + \dfrac{a^2 \Vone}{\adia'^2} \sonepert + \dfrac{a^2
  \Vtwo}{\adia'^2} \stwopert.
\end{equation}
We have also introduced a rotated 2-form quantity (with respect to
the field-manifold) from the potential according to
\begin{equation}
\left(
\begin{array}{ccc}
\vsep
\Vadiaadia & \Vadiaone & \Vadiatwo \\
\vsep
\Voneadia & \Veleven & \Vtwelve \\
\Vtwoadia & \Vtwentyone & \Vtwentytwo
\end{array}
\right) \negthickspace = \calM
\left(
\begin{array}{ccc}
  \vsep \dfrac{\Vall_{\matnd\matnd}}{\Afac^2} & \dfrac{\Vall_{\matnd
      \modP}}{\Afac \sqrt{\cplP}}
  & \dfrac{\Vall_{\matnd \modR}}{\Afac \sqrt{\cplR}} \\ \vsep
  \dfrac{\Vall_{\modP \matnd}}{\Afac \sqrt{\cplP}} &
  \dfrac{\Vall_{\modP \modP}}{\cplP} & \dfrac{\Vall_{\modP
      \modR}}{\sqrt{\cplP \cplR}} \\
  \dfrac{\Vall_{\modR \matnd }}{\Afac \sqrt{\cplR}} &
  \dfrac{\Vall_{\modR \modP }}{\sqrt{\cplR \cplP }} &
  \dfrac{\Vall_{\modR \modR}}{\cplR}
\end{array}
\right)
\calM^{-1},
\end{equation}
ensuring the symmetry properties
\begin{equation}
\Vadiaone = \Voneadia, \quad \Vadiatwo=\Vtwoadia, \quad
\Vtwentyone=\Vtwelve.
\end{equation}
Note that the special form of the effective potential in
Eq.~(\ref{eq:potall}) has been used to perform some simplifications
\begin{align}
  \Vadiaone &=\dfrac{1}{\cos 2\thetaone} \left[4
    \left(\dfrac{\alphafac_\modP}{\sqrt{\cplP}} \cos \thetatwo +
      \dfrac{\alphafac_\modR}{\sqrt{\cplR}} \sin \thetatwo \right)
    \left(-\sin \thetaone \Vone + \cos \thetaone \Vadia \right) +
    \dfrac{\sin 2 \thetaone}{2} \left(\Veleven - \Vadiaadia \right)
  \right],
\end{align}
and
\begin{align}
  \Vadiatwo &=\dfrac{1}{\cos \thetaone} \left[4
    \left(-\dfrac{\alphafac_\modP}{\sqrt{\cplP}} \sin \thetatwo +
      \dfrac{ \alphafac_\modR}{\sqrt{\cplR}} \cos \thetatwo \right)
    \left(-\sin \thetaone \Vone  + \cos
      \thetaone \Vadia \right) + \sin \thetaone \Vtwelve \right].
\end{align}
In Eq.~(\ref{eq:sonepertevol}), it was also convenient to introduce
the exact potential derivative
\begin{equation}
\begin{aligned}
  \Zeleven \equiv \dfrac{\partial \Vone}{\partial \sone} & = \Veleven
  + \cos \thetaone \sin \thetaone
  \left(\dfrac{\alphafac_\modP}{\sqrt{\cplP}} \cos \thetatwo +
    \dfrac{\alphafac_\modR}{\sqrt{\cplR}} \sin \thetatwo \right)
  \left(-\sin \thetaone \Vone + \cos \thetaone \Vadia \right).
\end{aligned}
\end{equation}
Similarly, for the other entropy mode one gets
\begin{equation}
\label{eq:stwopertevol}
\begin{aligned}
  \stwopert'' & + 2 \calH \stwopert' - \dfrac{2\adia'}{\tan \thetaone}
  \left(\dfrac{a^2 \Vtwo}{\adia'^2} +
    \dfrac{\alphafac_\modP}{\sqrt{\cplP}} \sin \thetatwo - \dfrac{
      \alphafac_\modR}{\sqrt{\cplR}} \cos \thetatwo \right) \sonepert'
  + \Bigg[-\Delta + a^2 \Vtwentytwo + \dfrac{2 a^2 \Vtwo}{\tan^2
    \thetaone} \left(-\dfrac{\alphafac_\modP}{\sqrt{\cplP}} \sin
    \thetatwo \right. \\ & + \left.
    \dfrac{\alphafac_\modR}{\sqrt{\cplR}} \cos \thetatwo \right) -
  \dfrac{a^4 \Vtwo^2}{\adia'^2 \sin^2 \thetaone} -\adia'^2 \cos^2
  \thetaone \left(\dfrac{\betafac_{\modP \modP}}{\cplP} \sin^2
    \thetatwo + \dfrac{\betafac_{\modR \modR}}{\cplR} \cos^2 \thetatwo
    - \dfrac{\betafac_{\modP\modR}}{\sqrt{\cplP \cplR}} \sin
    2\thetatwo\right) \\ & - \dfrac{1+\sin^2 \thetaone}{\tan^2
    \thetaone}\adia'^2 \left( - \dfrac{ \alphafac_\modP}{\sqrt{\cplP}}
    \sin \thetatwo + \dfrac{\alphafac_\modR}{\sqrt{\cplR}} \cos
    \thetatwo \right)^2 \Bigg] \stwopert + \Bigg\{ \left(\dfrac{-3 +
      \cos 2\thetaone}{\sin^2 \thetaone} a^2 \Vone + \dfrac{4}{\tan
      \thetaone} a^2 \Vadia \right) \\ & \times \left(
    \dfrac{\alphafac_\modP }{\sqrt{\cplP}}\sin \thetatwo -
    \dfrac{\alphafac_\modR}{\sqrt{\cplR}} \cos \thetatwo \right) +
  \dfrac{2a^2 \Vtwo}{\sin \thetaone \tan \thetaone} \left(
    \dfrac{\alphafac_\modP}{\sqrt{\cplP}} \cos \thetatwo +
    \dfrac{\alphafac_\modR}{\sqrt{\cplR}} \sin \thetatwo \right) \\ &
  - \dfrac{2 a^2 \Vtwo}{\adia'^2} \left(\dfrac{a^2 \Vone}{\sin^2
      \thetaone} + \dfrac{3 \calH \adia' + a^2 \Vadia}{\tan \thetaone}
  \right) - \dfrac{\sigma'^2}{\sin\thetaone \tan \thetaone} \left[
    \left(-\dfrac{\alphafac_\modP^2}{\cplP} +
      \dfrac{\alphafac_\modR^2}{\cplR} \right) \sin 2\thetatwo +
    \dfrac{2\alphafac_\modP \alphafac_\modR }{\sqrt{\cplP\cplR}} \cos
    2\thetatwo \right] \Bigg\} \sonepert \\ & = \dfrac{2 a^2
    \Vtwo}{\calH} \curvpert'.
\end{aligned}
\end{equation}
The equation of motion Eq.~(\ref{eq:curvpertevol}) for $\curvpert$ can
also be expressed in terms of the entropy modes only by
differentiation
\begin{equation}
\label{eq:curvpertevoldyn}
\begin{aligned}
\curvpert'' & - 2 \left(\calH + \dfrac{\calH'}{\calH} + \dfrac{a^2
  \Vadia}{\adia'} \right) \curvpert' - \Delta \curvpert = -\dfrac{2
  \calH}{\adia'} \left( \dfrac{a^2 \Vone}{\adia'} \sonepert' +
  \dfrac{a^2 \Vtwo}{\adia'} \stwopert'\right) - \dfrac{2
  \calH}{\adia'} \Bigg\{ \dfrac{\tan 2 \thetaone}{2} \left(a^2
  \Veleven - a^2 \Vadiaadia \right) \\ & - \left[ \sin
  \thetaone\left(\sin^2 \thetaone + \dfrac{4}{\cos 2\thetaone} \right)
  a^2 \Vone + \cos \thetaone \left(\cos^2 \thetaone - \dfrac{4}{\cos
  2 \thetaone} \right) a^2 \Vadia \right]
  \left(\dfrac{ \alphafac_\modP}{\sqrt{\cplP}} \cos
  \thetatwo + \dfrac{\alphafac_\modR}{\sqrt{\cplR}}
   \sin \thetatwo \right) \\ +
  &\dfrac{\cos^2 \thetaone}{\tan \thetaone} a^2 \Vtwo
  \left(-\dfrac{\alphafac_\modP}{\sqrt{\cplP}}  \sin
  \thetatwo + \dfrac{\alphafac_\modR}{\sqrt{\cplR}}
   \cos \thetatwo \right) - \dfrac{a^4
  \Vtwo^2}{\adia'^2 \tan\thetaone} + \dfrac{a^2
  \Vone}{\adia'^2} \left[a^2 \Vadia + \adia' \left(4 \calH -
  \dfrac{\calH'}{\calH} \right) \right] \Bigg\} \sonepert \\ & -
  \dfrac{2 \calH}{\adia'} \Bigg\{\tan \thetaone a^2 \Vtwelve +
  \left[ \dfrac{(-3 + \cos 2 \thetaone)^2}{2 \sin 2 \thetaone} a^2
  \Vone + \dfrac{\cos 2 \thetaone -7}{2} a^2 \Vadia \right]
  \left(\dfrac{\alphafac_\modP}{\sqrt{\cplP}}  \sin
  \thetatwo - \dfrac{\alphafac_\modR}{\sqrt{\cplR}}
   \cos \thetatwo \right) \\ & + \dfrac{a^2
  \Vtwo}{\adia'^2} \left(a^2 \Vadia + \dfrac{a^2 \Vone}{\tan
  \thetaone} + \adia' \left(4 \calH - \dfrac{\calH'}{\calH} \right)
  \right] \Bigg\} \stwopert.
\end{aligned}
\end{equation}
In order to set quantum initial conditions during the inflationary
eras, it is more convenient to use the Mukhanov
variable~\cite{Mukhanov:1990me} $\mukha = \adia' \curvpert/\calH$
whose evolution is readily obtained from
Eq.~(\ref{eq:curvpertevoldyn})
\begin{equation}
\label{eq:mukhaevol}
\begin{aligned}
  \mukha'' & + 2 \calH \mukha' + \Bigg\{-\Delta + a^2 \Vadiaadia -
  \left(\dfrac{\cos3 \thetaone - 5 \cos \thetaone}{4} a^2 \Vone +
    \cos^2 \thetaone \sin \thetaone a^2 \Vadia \right) \left(
    \dfrac{\alphafac_\modP}{\sqrt{\cplP}} \cos \thetatwo +
    \dfrac{\alphafac_\modR}{\sqrt{\cplR}} \sin \thetatwo \right) \\ &
  + a^2 \Vtwo \cos^2 \thetaone \left(-
    \dfrac{\alphafac_\modP}{\sqrt{\cplP}} \sin \thetatwo +
    \dfrac{\alphafac_\modR}{\sqrt{\cplR}} \cos \thetatwo \right) -
  a^4\dfrac{\Vone^2 + \Vtwo^2}{\adia'^2} + \dfrac{a^2 \Vadia}{\adia'}
  \left(4 \calH - 4 \dfrac{\calH'}{\calH}\right) + 2 \left(\calH -
    \dfrac{\calH'}{\calH} \right) \\ & \times \left(2\calH +
    \dfrac{\calH'}{\calH} \right) \Bigg\} \mukha = -2 \left(
    \dfrac{a^2 \Vone}{\adia'} \sonepert' + \dfrac{a^2 \Vtwo}{\adia'}
    \stwopert'\right) - 2 \Bigg\{ \dfrac{\tan 2 \thetaone}{2}
  \left(a^2 \Veleven - a^2 \Vadiaadia \right) \\ & - \left[ \sin
    \thetaone\left(\sin^2 \thetaone + \dfrac{4}{\cos 2\thetaone}
    \right) a^2 \Vone + \cos \thetaone \left(\cos^2 \thetaone -
      \dfrac{4}{\cos 2 \thetaone} \right) a^2 \Vadia \right]
  \left(\dfrac{\alphafac_\modP}{\sqrt{\cplP}}  \cos \thetatwo +
    \dfrac{\alphafac_\modR}{\sqrt{\cplR}}  \sin \thetatwo
  \right) \\ + &\dfrac{\cos^2 \thetaone}{\tan \thetaone} a^2 \Vtwo
  \left(-\dfrac{\alphafac_\modP}{\sqrt{\cplP}}  \sin \thetatwo +
    \dfrac{\alphafac_\modR}{\sqrt{\cplR}} \cos \thetatwo
  \right) - \dfrac{a^4 \Vtwo^2}{\adia'^2 \tan\thetaone} + \dfrac{a^2
    \Vone}{\adia'^2} \left[a^2 \Vadia + \adia' \left(4 \calH -
      \dfrac{\calH'}{\calH} \right) \right] \Bigg\} \sonepert \\ & - 2
  \Bigg\{\tan \thetaone a^2 \Vtwelve + \left[ \dfrac{(-3 + \cos 2
      \thetaone)^2}{2 \sin 2 \thetaone} a^2 \Vone + \dfrac{\cos 2
      \thetaone -7}{2} a^2 \Vadia \right]
  \left(\dfrac{\alphafac_\modP}{\sqrt{\cplP}}  \sin \thetatwo -
    \dfrac{\alphafac_\modR}{\sqrt{\cplR}}  \cos \thetatwo
  \right) \\ & + \dfrac{a^2 \Vtwo}{\adia'^2} \left(a^2 \Vadia +
    \dfrac{a^2 \Vone}{\tan \thetaone} + \adia' \left(4 \calH -
      \dfrac{\calH'}{\calH} \right) \right] \Bigg\} \stwopert.
\end{aligned}
\end{equation}
\end{widetext}
The equations (\ref{eq:sonepertevol}), (\ref{eq:stwopertevol}) and
(\ref{eq:mukhaevol}) form a closed system of equations. Indeed the
source terms involving the comoving curvature perturbation $\curvpert$
in Eqs.~(\ref{eq:sonepertevol}) and (\ref{eq:stwopertevol}) can be
expressed in terms of $\mukha$ as
\begin{equation}
\label{eq:curvpertsource}
2\dfrac{a^2 \Vall}{\calH} \curvpert' = 2 \dfrac{a^2 \Vall}{\adia'}
\mukha' +  2 \dfrac{a^2 \Vall}{\adia'^2} \left[a^2 \Vadia + \adia'\left(
    2 \calH + \dfrac{\calH'}{\calH} \right) \right] \mukha.
\end{equation}
One can check that Eqs.~(\ref{eq:sonepertevol}),
(\ref{eq:curvpertevoldyn}) and (\ref{eq:mukhaevol}) match with the
ones obtained for two fields in
Refs.~\cite{DiMarco:2002eb,Ashcroft:2003xx} for $\modRpert = \modR=0$,
whereas Eq.~(\ref{eq:stwopertevol}) decouples and admits $\stwopert=0$
as a solution, up to transformation $\theta \rightarrow \pi/2 -
\thetaone$. As shown in Sect.~\ref{sect:background}, since the dilaton
$\modR$ relaxes toward $\modR=0$ rapidly, we expect to recover a two
field like behavior also at the perturbation level. In the next
sections we numerically solve the system of equations
(\ref{eq:sonepertevol}), (\ref{eq:stwopertevol}) and
(\ref{eq:mukhaevol}) to compute the resulting power spectra at the end
of inflation.

\subsubsection{Numerical results}
\label{sect:linpert_numres}
\begin{figure}
\begin{center}
\includegraphics[width=8.5cm]{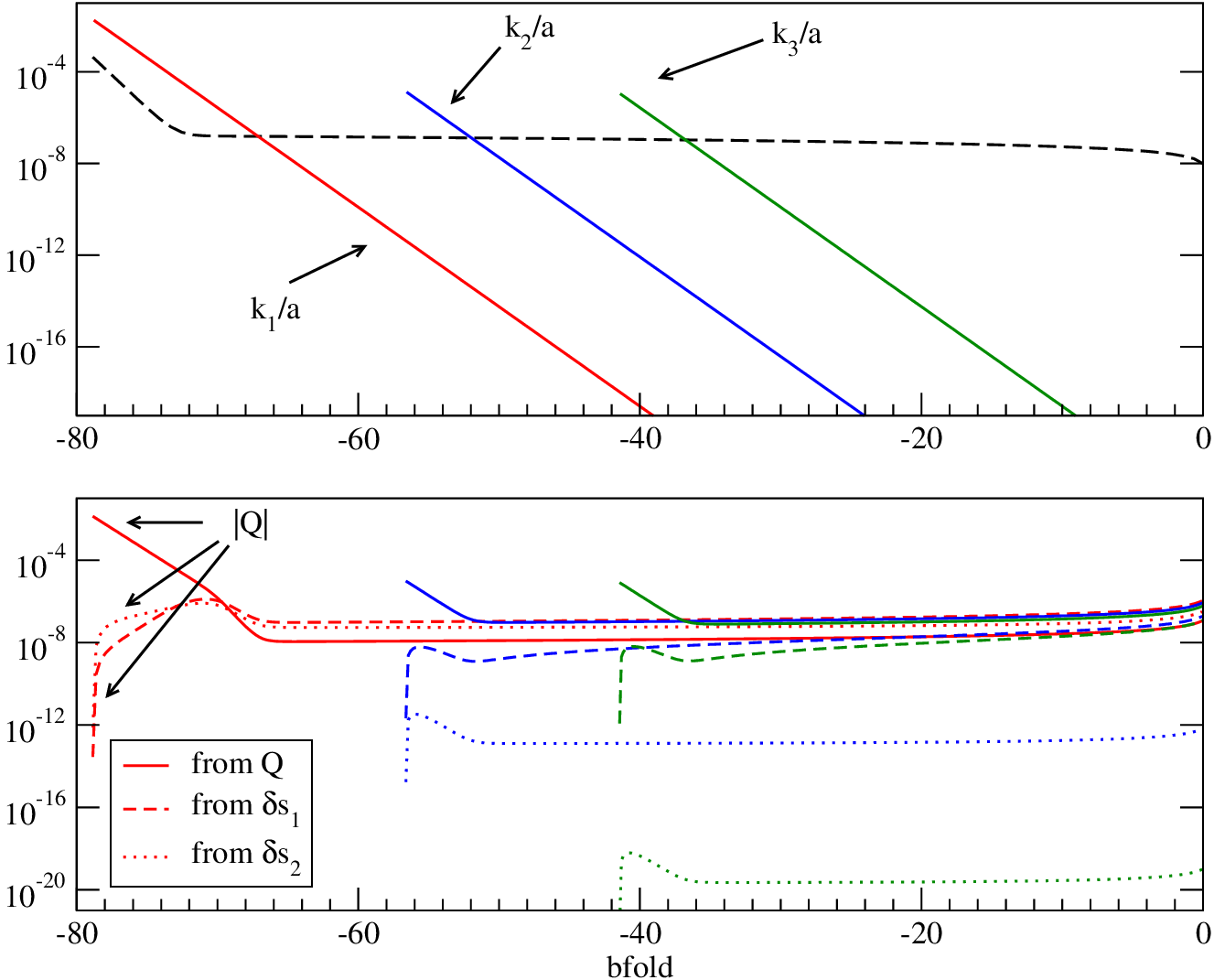}
\includegraphics[width=8.5cm]{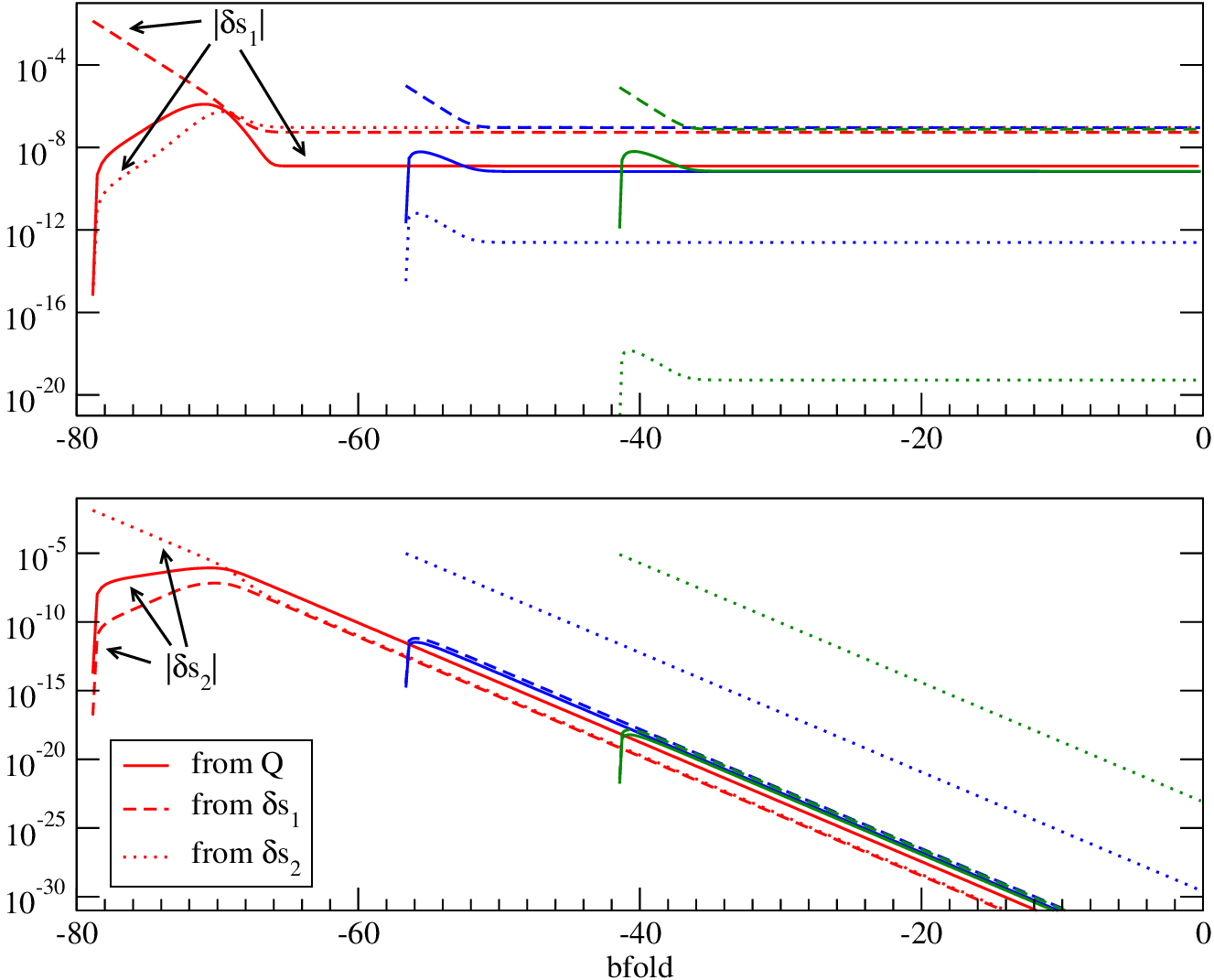}
\caption{Evolution of the adiabatic and entropy perturbations in bfold
  time (efold before the end of inflation), for three wavenumbers
  $\kwav_\one < \kwav_\two < \kwav_\three$. In the top frame, the
  Hubble parameter and the physical wavenumbers $\kwav/a$ are plotted.
  The next three frames shows the evolution of each $\mukha$,
  $\sonepert$ and $\stwopert$ modes, respectively, obtained from the
  three independent initial conditions where only one of the mode is
  in a Bunch-Davies vacuum and the others vanish. Only the wavenumber
  leaving the Hubble radius close to the background $\modR$ dominated
  expansion exhibits strong couplings between the adiabatic and
  entropy modes. The background model parameters are the same as in
  Fig.~\ref{fig:2phases} and we have chosen $\baremassnd=10^{-6}$.}
\label{fig:pertsketch}
\end{center}
\end{figure}

\begin{figure}
\begin{center}
\includegraphics[width=8.5cm]{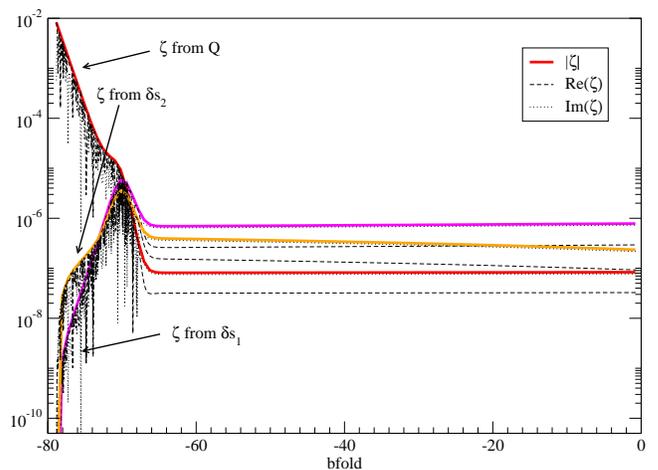}
\caption{Real, imaginary and modulus of the comoving curvature
  perturbation mode $\curvpert = \mukha \calH/\adia' =
  \mukha/\dot{\adia}$ corresponding to the $\kwav_\one$ wavenumber,
  and model parameters, of Fig.~\ref{fig:pertsketch}. The mode behaves
  as a free wave under the Hubble radius whereas it is damped after
  Hubble exit while sourced by the entropy perturbations. Note also
  the change of slope of the modulus during the background transition
  between the $\modR$ dominated expansion and the subsequent
  inflationary era (see Sect.~\ref{sect:background}).}
\label{fig:zetaosci}
\end{center}
\end{figure}

For a given background evolution (see Sect.~\ref{sect:background}),
and in the Fourier space with respect to the comoving spatial
coordinates, the solutions of Eqs.~(\ref{eq:sonepertevol}),
(\ref{eq:curvpertevoldyn}) and (\ref{eq:mukhaevol}) are uniquely
determined by the initial values of $\sonepert$, $\stwopert$, $\mukha$
and their first order time derivative. Motivated by the quantum origin
picture of the cosmological perturbations, in the Einstein Frame the
scalar and tensor modes are decoupled and we can treat $\sonepert$,
$\stwopert$ and $\mukha$ as independent stochastic variables deep
inside the Hubble radius~\cite{Mukhanov:1990me,
  Salopek:1988qh,Tsujikawa:2002qx, Martin:2004um}.  Following the
method of Ref.~\cite{Tsujikawa:2002qx}, since we are interested in the
two-point correlation function between the different rotated fields
$\sonepert$, $\stwopert$ and $\mukha$ (or $\curvpert$), we will
consider only the solutions obtained by starting with one mode in the
Bunch-Davies vacuum while the others vanish, and the other mode
permutations. In addition to providing enough information to compute all
the power spectra, they clearly exhibit the sourcing effects of one
field from the others.  For numerical convenience, each Fourier mode
$\kwav$ will be assumed to appear at a given time $\eta_\uquant$ such
that
\begin{equation}
\label{eq:initime}
\dfrac{\kwav}{\calH(\eta_\uquant)} = \constdec,
\end{equation}
where $\kwav$ is the comoving wave number and $\constdec$ a constant
verifying $\constdec \gg 1$ and characterizing the decoupling
limit~\cite{Salopek:1988qh,Ashcroft:2003xx}. For a Bunch-Davies vacuum,
the value of $\constdec$ does not usually change the power spectra,
provided it is big enough to allow free wave solution\footnote{This is
  no longer true for trans-Planckian initial
  conditions~\cite{Martin:2003sg, Easther:2005yr}.}. For $\qmodeS= a
\sqrt{2} \, \kwav^{3/2}\mukha$ (and respectively $\qmodeS= a \sqrt{2}
\, \kwav^{3/2}\sonepert$, $\qmodeS= a \sqrt{2} \,
\kwav^{3/2}\stwopert$), we therefore set at
$\eta=\eta_\uquant$~\cite{Kaloper:2002cs, Martin:2003kp}
\begin{equation}
\label{eq:qmodeini}
\begin{aligned}
\qmodeS = - \kappa \kwav, \quad
\dfrac{\qmodeS'}{\kwav} = i \kappa \kwav.
\end{aligned}
\end{equation}
In Fig.~\ref{fig:pertsketch} the numerical solutions of
Eqs.~(\ref{eq:sonepertevol}) to (\ref{eq:mukhaevol}) have been plotted
for the same background evolution as in Fig.~\ref{fig:2phases} and for
three different values of the comoving wave number $\kwav$ with
$\kwav_\one < \kwav_\two < \kwav_\three$. In the top frame, the
evolution of the Hubble parameter and the three physical wavenumbers
$\kwav/a$ are plotted as a function of the number of efold before the
end of inflation (``bfold'' in the following). A subtlety occurs
however for the ``very large'' wavelength modes due to the existence
of a non-inflationary era when the background evolution is dominated
by the rolling of the $\modR$ field (see Sect.~\ref{sect:background}
and Fig.~\ref{fig:2phases}). Indeed, if the initial conditions in the
five-dimensional setup are such that the initial value of $\modR$ is
far from vanishing, then there are some perturbation modes which
initially are super-Hubble and for which setting quantum initial
conditions does not make sense. However, since we are interested in
enough efolding to solve the flatness and homogeneity problems, such
large wavelength modes are generically still super-Hubble today and
thus non-observable. On the other hand, one cannot exclude that for a
small number of efold, let's say $60$ for instance, some of the large
wavelength perturbation modes entering the Hubble radius today may
have been created during the transition period between the $\modR$
dominated phase and the inflationary one, as $\kwav_\one$ in
Fig.~\ref{fig:pertsketch}. Such a mode starts initially under the
Hubble radius, but there is a maximum value of $\constdec$ for which
Eq.~(\ref{eq:initime}) has not solution (because $\eta_\uquant$ would
be too small). Our prescription is to discard the models for which
observable perturbation modes today would have been created on Hubble,
super-Hubble and sub-Hubble scales with $\constdec < 100$ initially.
{}From the second to the bottom frame in
Fig.~\ref{fig:pertsketch}, the evolution of $|\mukha(\bfold)|$,
$|\sonepert(\bfold)|$ and $|\stwopert(\bfold)|$ are plotted,
respectively. The straight, dashed and dot lines correspond to the
three independent initial conditions for the modes, $\mukha$ in a
Bunch-Davies vacuum and $\sonepert=\sonepert'=\stwopert=\stwopert'=0$,
$\sonepert$ in a Bunch-Davies vacuum and $\mukha=\mukha' = \stwopert =
\stwopert' = 0$, \dots, respectively.  Moreover, in each frame we have
plotted the three wavelength modes $\kwav_\one$, $\kwav_\two$ and
$\kwav_\three$ from the left to the right. In Fig.~\ref{fig:zetaosci},
we have plotted the real and imaginary parts of the comoving curvature
perturbation $\curvpert = \mukha \calH/\adia'$ for the mode
$\kwav_\one$ associated with the $\mukha$ solution of
Fig.~\ref{fig:pertsketch} and the background evolution of
Fig.~\ref{fig:2phases}.
This plot shows the transition between the oscillatory behavior when
the mode is under the Hubble radius and its damping on super-Hubble
scales. However, since we are precisely in a multifields case, even on
super-Hubble scales the comoving curvature perturbation slightly
evolves sourced by the entropy modes till the end of inflation. Also,
the rapid change in behavior of the Hubble parameter during the
transition between the $\modR$ dominated era and the subsequent
inflationary eras induces also a change of slope in the evolution of
all the perturbation modes (see Fig.~\ref{fig:pertsketch} and
Fig.~\ref{fig:zetaosci} around $\efold - \efold_\uend \simeq -70$).
This can be understood by first noting that before the transition, as
long as $\epsone > 1$, the physical wavelength of the mode goes deeper
into the Hubble radius with time whereas as soon as $\epsone < 1$ the
wavelength becomes closer and leaves the Hubble radius. Moreover the
amplitude of the sourcing effect of the entropy modes on the comoving
curvature perturbation is enhanced. As can be seen in
Fig.~\ref{fig:pertsketch} and Fig.~\ref{fig:zetaosci}, the adiabatic
mode is mainly produced by the entropy modes for $\kwav=\kwav_\one$
whereas this effect progressively disappears for the bigger
wavenumbers $\kwav_\two$ and $\kwav_\three$.  This comes from the fact
that only the $\kwav_\one$ mode evolves and leaves the Hubble radius
in a regime where the $\modR$ field is not yet vanishing (see
Fig.~\ref{fig:2phases}). Note that in Eqs.~(\ref{eq:sonepertevol}) to
(\ref{eq:stwopertevol}), the Hubble parameter appears also in the
effective potential of the modes. Its rapid (exponential) evolution
during the relaxation of $\modR$ enhances the coupling between the
entropy and adiabatic modes.  Also, at the transition, the background
fields are no longer in a stationary regime as discussed in
Sect.~\ref{sect:background}, and especially one expects $\thetaone'$
and $\thetatwo'$ to increase and reinforces the coupling between the
adiabatic and entropy modes. This effect is similar to the one
observed for double-inflation models~\cite{Tsujikawa:2002qx}.  For the
modes which evolve out of this stage, the $\modR$ field and its
perturbations are negligible, \ie $\thetatwo \simeq 0$ and the second
entropy mode $\stwopert$ decouples and decreases exponentially [see
  Eqs.~(\ref{eq:oldperttonewpert}), (\ref{eq:stwopertevol}) and the
  bottom frame in Fig.~\ref{fig:pertsketch}]. The dynamics is close to
a two fields like regime where the friction dominated evolution leads
to a weakly coupled entropy mode $\sonepert$~\cite{Tsujikawa:2002qx,
  Ashcroft:2003xx}. As can been seen in Fig.~\ref{fig:pertsketch},
after Hubble exit, for the wavenumbers $\kwav_\two$ and
$\kwav_\three$, the entropy perturbations $\sonepert$ are quasi-frozen
as it would be for a light scalar field.

\subsection{Tensor modes}

\begin{figure}
\begin{center}
\includegraphics[width=8.5cm]{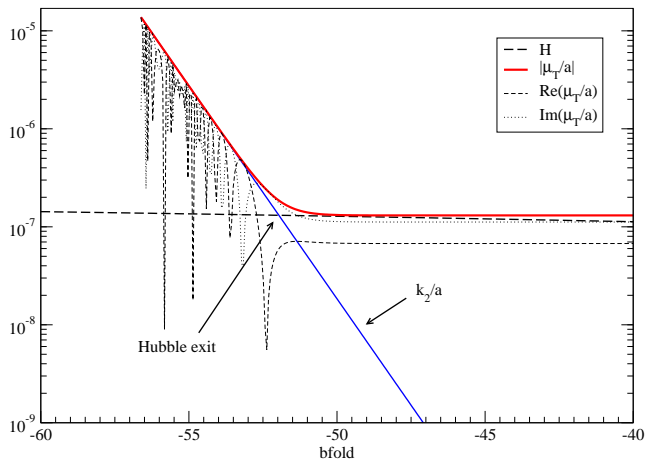}
\caption{Real, imaginary and modulus of the tensor perturbation mode
  corresponding to the $\kwav_\two$ wavenumber. The Hubble parameter
  is also shown and we recover the standard result $|\qmodeT/a| \simeq
  \hubble_\kwav$ with $\hubble_\kwav$ its value at Hubble exit. Note
  however that it takes few more efolds for the mode to remain
  constant which means that this approximation is valid only for a
  slow varying Hubble parameter (as during the slow-roll stage).}
\label{fig:tenssketch}
\end{center}
\end{figure}

In the Einstein frame, the scalar and tensor degrees of freedom are
decoupled. Therefore the equation of evolution for the tensor modes
remains the same as in General Relativity. For a flat perturbed FLRW
metric
\begin{equation}
\label{eq:perttensmetric}
\ud s^2 = - a^2 \ud \eta^2 + a^2 \left(\delta_{ij} + \tens_{ij}\right) \ud
x^i \ud x^j,
\end{equation}
where $\tens_{ij}$ is a traceless and divergenceless tensor
\begin{equation}
\delta^{ij} \tens_{ij} = \delta^{ik} \partial_k \tens_{ij} = 0,
\end{equation}
one gets~\cite{Mukhanov:1990me,Liddle:1993fq}
\begin{equation}
\label{eq:tensevol}
\tens_{ij}'' + 2 \calH \tens_{ij}' - \Delta \tens_{ij}=0.
\end{equation}
This equation can be numerically solved for each polarization state
$\tens$ by setting the initial quantum tensor modes
$\qmodeT=\kwav^{3/2} a \tens$ in the Bunch-Davies
vacuum [see
Eq.~(\ref{eq:qmodeini})]. In Fig.~\ref{fig:tenssketch} their evolution
has been plotted for wavenumber $\kwav_\two$ together with the
evolution of the Hubble parameter. As expected for a massless field,
the gravitational waves freeze on super-Hubble scales with
$|\qmodeT/a| \simeq \hubble_{\kwav}$ where $\hubble_\kwav$ is the
Hubble parameter at Hubble exit. Note however that the freezing occurs
a few efolds after Hubble exit.

\subsection{Primordial scalar and tensor power spectra}
\label{sect:linpert_power}

{}From the scalar and tensor modes evolution obtained in the previous
sections, the primordial power spectra are readily obtained from the
values taken by the adiabatic and entropy perturbations at the end of
inflation. In the Fourier space and Einstein frame, the scalar power
spectra are computed according to the method used in
Ref.~\cite{Tsujikawa:2002qx}, \ie
\begin{equation}
\label{eq:scalpowerspect}
\begin{aligned}
  \power_{ab} & = \dfrac{\kwav^3}{2 \pi^2} \sum_m
  \left[\obspert_m^{a}(\kwav)\right]^* \left[\obspert_m^{b}(\kwav)\right],
\end{aligned}
\end{equation}
where the observable perturbation modes $\obspert^a$ stands for
$\curvpert$, $\sonepert/\dot{\adia}$ and $\stwopert/\dot{\adia}$,
whereas the index ``$m$'' refers to the three independent quantum
initial conditions. The tensor power spectrum reads
\begin{equation}
\label{eq:tenspowerspect}
\begin{aligned}
  \powerTens(\kwav) & = \dfrac{2 \kwav^3}{\pi^2} \left| \tens(\kwav)
  \right|^2,
\end{aligned}
\end{equation}
where the polarization degrees of freedom have been included.

\begin{figure}
\begin{center}
\includegraphics[width=8.5cm]{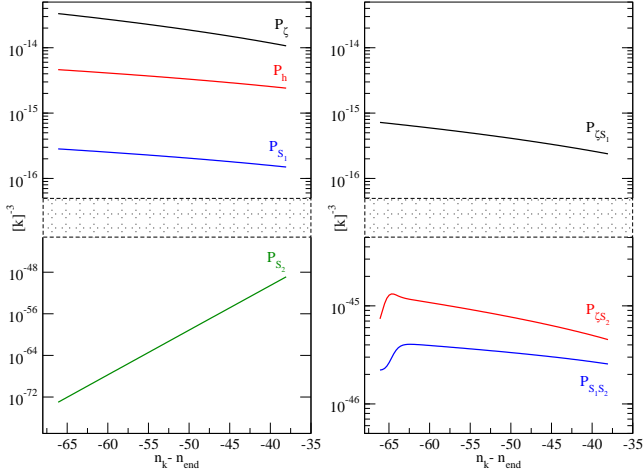}
\caption{Typical power spectra at the end of inflation for the
  perturbation modes of Fig.~\ref{fig:pertsketch} and
  Fig.~\ref{fig:tenssketch} which becomes super-Hubble after
  $\efold_\kwav$ efolds. The entropy power spectra are sub-dominant
  compared to the adiabatic one.}
\label{fig:powerspec}
\end{center}
\end{figure}

In Fig.~\ref{fig:powerspec}, we have plotted the typical power spectra
associated with the solutions computed in
Sect.~\ref{sect:linpert_numres}. The adiabatic power spectra
$\powerCurv$ is dominant compared to the two entropy ones $\powerSone$
and $\powerStwo$. Furthermore, the second entropy power spectrum
$\powerStwo$ is strongly damped and blue tilted compared to
$\powerSone$. In fact, this behavior is expected since the second
entropy mode $\stwopert$ follows the $\modR$ evolution during
inflation and vanishes exponentially with the total number of
efolds. As a result, the earlier the $\stwopert$ modes cross the
Hubble radius, the longer they sustain an exponential decay. Their
power spectrum at the end of inflation is blue tilted since the
smaller scales are accordingly less damped. However, since the
observable perturbation modes are expected to cross the Hubble radius
around $60$ efolds before the end of inflation, the entropy power
spectrum $\power_{\Stwo}$ cannot be significant at the end of
inflation. In particular, the expected cross-correlations for
$\kwav_\one$-like modes discussed in Sect.~\ref{sect:linpert_numres}
are also washed out by such a damping. They can still be seen in
Fig.~\ref{fig:powerspec} as a bump in the cross-correlation power
spectra $\power_{\zeta \Stwo}$ and $\power_{\Sone\Stwo}$, but with an
extremely small amplitude. It is important to recall that the
domination of $\modR$ during inflation has to occur before the last
$60$ efolds of inflation since it drives a non-accelerated expansion
in our model. As seen in Sect.~\ref{sect:background}, this comes from
the fact that $\cplR+\cplP=6$ whereas in a more generic scalar-tensor
model with $\cplR+\cplP<9/2$ the transition from $\modR$-dominated to
$\modP$-dominated inflation could be in the observable range and may
generate a significant power spectrum for the second entropy
mode~\cite{Adams:1997de, Kaloper:2003nv, Hunt:2004vt}. Nevertheless,
we will accordingly be focused in the following on the present
boundary inflation model where only the adiabatic, tensor and first
entropy modes are of interest for the CMB~\cite{Ashcroft:2003xx}.

The hierarchy in the power spectra amplitudes can be qualitatively
understood by noting that the entropy mode $\sonepert$ behaves
almost like a free massless field for moderate  values of the
coupling constant $\cplP$. Indeed, as can be seen in
Fig.~\ref{fig:pertsketch}, $\sonepert$ is weakly coupled to the
adiabatic mode and remains almost frozen once it becomes
super-Hubble. Neglecting the sourcing effects from the adiabatic
modes, as well as the variation of the Hubble parameter during the
few efolds after Hubble exit, one gets at the end of inflation
\begin{equation}
  \kwav^{3/2} \left| \sonepert \right|_\uend \simeq
  \dfrac{\hubblend_{\kwav}}{\sqrt{2}},
\end{equation}
where $\hubblend_{\kwav}$ is the dimensionless Hubble parameter
$\kappa \hubble$ at Hubble exit for each mode of wavenumber $\kwav$.
{}From Eqs.~(\ref{eq:epsoneVel}) and (\ref{eq:scalpowerspect}), keeping
in mind that $\epsone \simeq 1$ at the end of inflation, the resulting
power spectrum varies as
\begin{equation}
  \label{eq:powersoneapprox}
  \powerSone \simeq \dfrac{\hubblend_{\kwav}^2}{8\pi^2}
  \simeq \dfrac{1}{16} \power_h.
\end{equation}
Similarly, assuming that the adiabatic mode $\mukha$ gets massless
field-like fluctuations at Hubble exit, in the weakly coupled regime
the comoving curvature perturbation remains constant afterwards and
the adiabatic power spectrum can be approximated by
\begin{equation}
  \label{eq:powerzetaapprox}
  \powerCurv \simeq \dfrac{1}{8 \pi^2}
  \dfrac{\hubblend_{\kwav}^2} {\epsone_{\kwav}},
\end{equation}
where $\epsone_{\kwav}$ is the first slow-roll parameter at Hubble
exit.

One can check in Fig.~\ref{fig:powerspec} that
Eqs.~(\ref{eq:powersoneapprox}) and (\ref{eq:powerzetaapprox}) provide
a good approximation of the relative power spectra
amplitudes. Moreover, these equations show that the generated power
spectra do not depend significantly on the initial values of the
background fields which drive inflation. Indeed, as thoroughly
discussed in Sect.~\ref{sect:background}, the fields are rapidly
attracted toward a friction dominated regime during which their
evolution does no longer depend on their initial value. Obviously, the
initial conditions determine the total number of efolds, but at a
given efold before the end of inflation one may not expect
$\epsone_{\kwav}$ to be strongly influenced [see
  Eq.~(\ref{eq:epsoneVel})]. However, the amplitude of the
perturbations at Hubble exit directly depends on the Hubble parameter
value which involves both the bare mass $\baremassnd$ of the matter
field and the conformal factor $\Afac$. In order to disentangle their
respective effects, it is more convenient to consider the effective
mass of the matter field at the end of inflation
\begin{equation}
\label{effectivemattermass}
\baremasseff \equiv \Afac_\uend \baremassnd = \Afac_\uend \kappa \baremass.
\end{equation}
\begin{figure}
\begin{center}
\includegraphics[width=8.5cm]{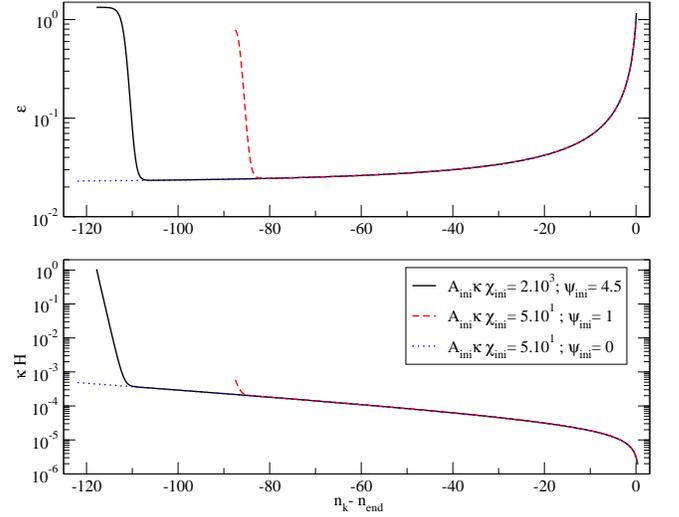}
\caption{Evolution of the first slow-roll parameter $\epsone$ and the
  dimensionless Hubble parameter $\hubblend$ for different initial
  values of the background fields. The $\modP$ coupling constant has
  been set to $\cplP=0.1$ and the effective mass of the matter field
  at the end of inflation is $\baremasseff=6.2\times 10^{-6}$. The
  existence of a friction dominated regime for the background fields
  leads to observable power spectra which are insensitive to the
  initial values of the background fields.}
\label{fig:otherbackevol}
\end{center}
\end{figure}
In Fig.~\ref{fig:otherbackevol}, the slow-roll and Hubble parameters
have been plotted for several arbitrarily chosen value of $\Afac_\uini
\matnd_\uini$ and $\modR_\uini$, the other parameters $\cplP$ and
$\baremasseff$ being fixed. As the plots show, the last efolds of
evolution cannot be differentiated and we have checked by a direct
computation that this is also the case for the corresponding power
spectra at the end of inflation. In fact, the only effects
$\Afac_\uini \matnd_\uini$ and $\modR_\uini$ might induce concern the
transition at the end of the $\modR$-dominated expansion, which is, as
explained before, hardly observable for the CMB.

The remaining degrees of freedom in the power spectra are the
effective mass $\baremasseff$ and the moduli coupling constant
$\cplP$. These parameters are expected to have significant observable
effects. On one hand, the effective mass fixes the overall amplitude
of the Hubble parameter and is therefore directly related to the
amplitude of the primordial perturbations. On the other hand, the
coupling constant $\cplP$ encodes how much the conformal factor may
run during the generation of the observable perturbations. From
Eq.~(\ref{eq:potall}), such a running renders the potential steeper
and one may expect bigger tilts for the power spectra.
\begin{figure}
\begin{center}
\includegraphics[width=8.5cm]{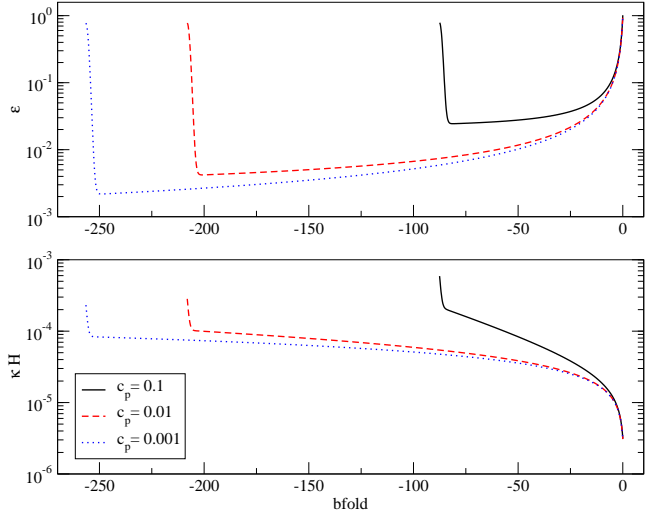}
\caption{Evolution of the first slow-roll parameter $\epsone$ and the
  dimensionless Hubble parameter $\hubblend$ for $\cplP=0.1$,
  $\cplP=0.01$ and $\cplP=0.001$. The effective matter field mass is
  $\baremasseff=6.2\times 10^{-6}$. The running of the conformal
  factor during the generation of the observable perturbations leads
  to more tilted power spectra (see Fig.~\ref{fig:powerevol}).}
\label{fig:backevol}
\end{center}
\end{figure}
\begin{figure}
\begin{center}
\includegraphics[width=8.5cm]{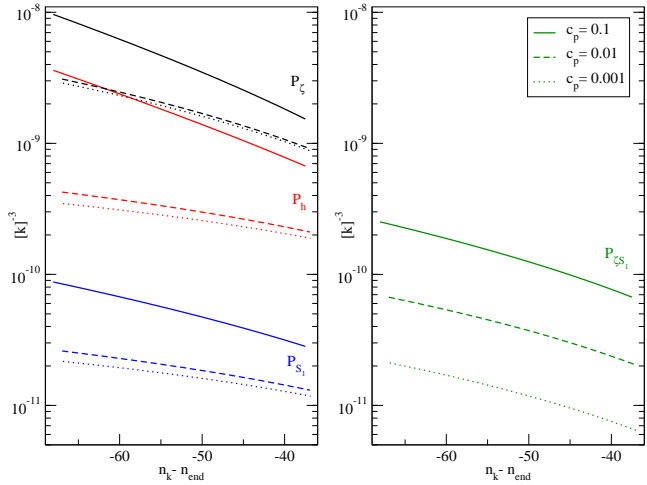}
\caption{Dependence of the power spectra at the end of inflation with
  respect to the moduli coupling constant $\cplP$. Larger slopes are
  obtained for higher values of the coupling constant, as expected
  from the running of the conformal factor during the generation of
  the perturbation modes (see also Fig.~\ref{fig:backevol}).}
\label{fig:powerevol}
\end{center}
\end{figure}

In Fig.~\ref{fig:backevol}, the slow-roll and Hubble parameters have
been plotted for several values of $\cplP$, the effective mass at the
end of inflation being fixed. The slopes of these two functions are
effectively steeper for larger values of the coupling constant and the
resulting power spectra may be qualitatively guessed from
(\ref{eq:powersoneapprox}) and (\ref{eq:powerzetaapprox}). The direct
computations of the power spectra at the end of inflation have been
plotted in Fig.~\ref{fig:powerevol} and confirm this behavior.

\begin{figure}
\begin{center}
\includegraphics[width=8.5cm]{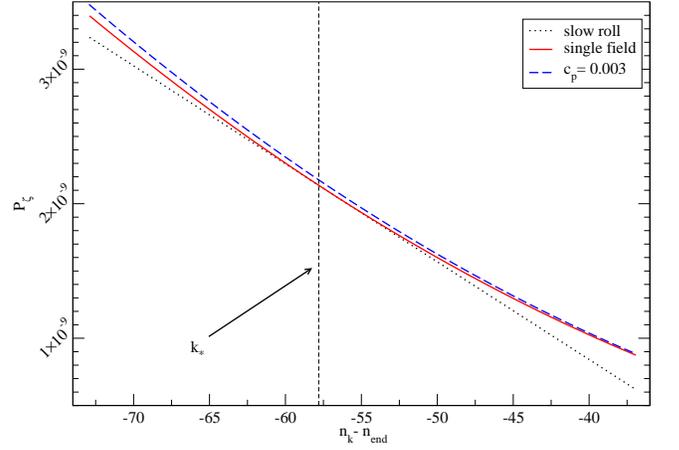}
\caption{The adiabatic power spectrum at the end of a weakly coupled
  boundary inflation compared to the one obtained in single field
  chaotic inflation. The dotted line is the first order slow-roll
  approximation of the chaotic model spectrum around the pivot scale
  $\kpivot$. The deviations between the weakly coupled boundary and
  chaotic models come from the slight running of the conformal
  factor.}
\label{fig:compare}
\end{center}
\end{figure}

In order to dress a qualitative understanding of the above-mentioned
effects, we have compared in Fig.~\ref{fig:compare} the adiabatic
power spectrum generated during a weakly coupled boundary inflation
with $\cplP=3\times 10^{-3}$ with the one associated with the standard
single field chaotic inflation ($\cplP=0$). Moreover, we have plotted
the first order slow-roll approximation of the chaotic model around a
pivot scale $\kpivot$. In this limit, the adiabatic power spectrum
simplifies to~\cite{Martin:1999wa}
\begin{equation}
\label{eq:slowrollpowerzeta}
\powerCurv \underset{\usr}{=} \dfrac{\hubblend^2}{8 \pi^2\epsone }
\left[ 1 - 2 \left(\constsr+1
  \right) \epsone - \constsr \epstwo - \left(2\epsone - \epstwo\right) \ln
  \dfrac{\kwav}{\kpivot} \right],
\end{equation}
where all the parameters are evaluated at the pivot scale. The
constant $\constsr\simeq -0.73$ and $\epstwo \equiv\ud \ln \epsone / \ud
\efold$ is the second slow-roll parameter. The deviations observed in
Fig.~\ref{fig:compare} between the slow-roll and single field power
spectra come from the natural running of the spectral index which is
not grabbed by the first order slow-roll approximation. The
differences between the chaotic and boundary spectra are due to the
slight running of the conformal factor obtained for $\cplP=3 \times
10^{-3}$. More qualitatively, for non-vanishing values of the moduli
coupling constant, the effective mass of the matter field is not
constant during the generation of the observable perturbations and
induces deviations with respect to a single field model.

\begin{figure}
\begin{center}
\includegraphics[width=8.5cm]{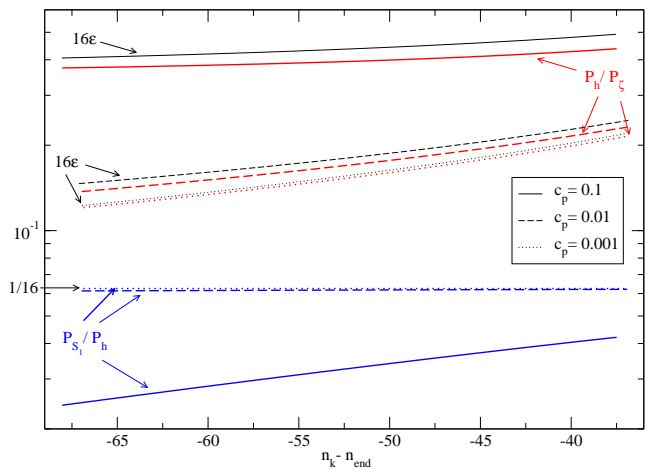}
\caption{Ratios between the power spectra of Fig.~\ref{fig:powerevol}
  and their evolution with respect to the moduli coupling constant
  $\cplP$. The ``consistency check'' $\powerTens/\powerCurv = 16
  \epsone$ is recovered in the weak coupling limit while significant
  deviations appear for $\cplP=0.1$ due to the sourcing effects
  between the modes after Hubble exit.}
\label{fig:ratioevol}
\end{center}
\end{figure}

As previously pointed, Eqs.~(\ref{eq:powersoneapprox}) and
(\ref{eq:powerzetaapprox}) are relevant only if the modes do not
evolve significantly after Hubble exit. This is obviously the case for
the single field model, but certainly no longer true for high values
of the moduli coupling constant. In Fig.~\ref{fig:ratioevol}, we have
plotted the ratio between the computed power spectra for several
values of $\cplP$. Some deviations show up for $\cplP>0.01$ in the
ratio tensor to scalar and in the entropy
modes~\cite{Starobinsky:1994mh,Garcia-Bellido:1995fz}.

In the next section, after setting up a toy cosmological model, we use
the above numerical method to derive the resulting CMB anisotropies
and perform a Monte Carlo Markov Chains exploration of the model
parameter space given the first year WMAP data. As expected from the
behaviour of the primordial power spectra, we find both the effective
mass of the matter field at the end of inflation and the moduli
coupling constant $\cplP$ to be constrained.

\section{CMB anisotropies}
\label{sect:cmb}

\subsection{Cosmological framework}
\label{sect:cmb_cosmoframe}

As mentioned in Sect.~\ref{sect:background}, in the brane world
picture, the moduli $\modR$ and $\modP$ are associated with the
position of the branes in the five-dimensional spacetime while the
field $\mat$ lives on the brane supposed to be our universe. As a
result, a natural setup is to consider that the moduli fields remain
in the late-time cosmology and in particular today whereas the
radiation, matter and dark matter sectors are sourced by the decay of
the field $\mat$ through a reheating period~\cite{Turner:1983he,
  Shtanov:1994ce, Kofman:1997yn} (see Fig.~\ref{fig:endinf}). In the
following we consider a toy cosmological background model by assuming
instantaneous reheating~\cite{Liddle:2003as}. In this respect, the
energy density at the beginning of the radiation era is completely
sourced by the energy density of the matter field $\mat$ at the end of
the inflationary period. This allows an estimation of the scale factor
at the end of inflation
\begin{equation}
  \label{eq:redshiftinf}
  \ln\dfrac{\aJF_\uend}{\aJF_\zero}  \simeq -\ln
  \dfrac{\aJF_\zero}{\aJF_\ueq} + \dfrac{1}{4}
  \ln\dfrac{\energyJF_\ueq} {\energyJF_\uend},
\end{equation}
where the index zero refers to today, ``eq'' at the equivalence
between radiation and matter, and with the energy density
$\energyJF_\uend = \energyJF_\mat$.

On the other hand, since we are dealing with a multiscalar-tensor
theory in the late-time cosmology, and especially today, the evolution
of the conformal factor and its first and second gradients are
constrained by various solar-system, astrophysical and cosmological
measurements~\cite{Damour:1998ae, Perrotta:1999am, Chen:1999qh,
  Baccigalupi:2000je, Amendola:2000ub, Riazuelo:2001mg,
  Acquaviva:2004ti, Esposito-Farese:2004tu,Uzan:2004qr,
  Schimd:2004nq}. As discussed in Sect.~\ref{sect:background}, the
field $\modR$ is rapidly driven toward zero and we will safely assume
in the following that it is indeed the case during the radiation and
matter eras. {}From Eq.~(\ref{eq:firstgrads}), the only non-vanishing
first conformal gradient is $\alphafac_\modP$, which remains constant
during the cosmological evolution. The strongest constraint comes from
the solar system~\cite{Schimd:2004nq, Davis:2005au}. From the Cassini
spacecraft measurements, one gets~\cite{Bertotti:2003rm}
\begin{equation}
\label{eq:firstgradtoday}
\metric^{ab} \alphafac_{a} \alphafac_{b}
\underset{\mathrm{solar}}{<} 5\times 10^{-7}
\quad \Rightarrow \quad \cplP \underset{\mathrm{solar}}{<} 2
\times 10^{-5} \,.
\end{equation}
Note that once $\modR=0$, the only non-vanishing second conformal
gradient $\betafac_{\modR\modR}$ is poorly constrained. This comes
from the fact that the post-Newtonian parameter $\gammaPN$ involves
only the combination $\alphafac^a \alphafac^b \betafac_{ab}$ which
vanishes in that case, whatever the value of
$\cplR$~\cite{Damour:1992we}. The other constraints coming from the
variation of the ``Cavendish'' gravitational
constant~\cite{Williams:1995nq}
\begin{equation}
\label{eq:kappacav}
\kappacav^2 = \kappa^2 \Afac^2 (1+2\metric^{ab}\alphafac_a \alphafac_b),
\end{equation}
are found to be satisfied once Eq.~(\ref{eq:firstgradtoday}) is. The
cosmological constraints on scalar-tensor gravity are less stringent
than the solar-system ones but allow one to probe the cosmic times.
They lead to the two-sigma upper bound~\cite{Damour:1998ae,
  Perrotta:1999am, Chen:1999qh, Amendola:2000ub,
  Baccigalupi:2000je,Riazuelo:2001mg, Rhodes:2003ev, Nagata:2003qn,
  Acquaviva:2004ti, Schimd:2004nq}
\begin{equation}
  \label{eq:firstgradcosmo}
  \metric^{ab} \alphafac_{a} \alphafac_{b}
  \underset{\mathrm{cosmo}}{<} 2 \times 10^{-3} \quad \Rightarrow
  \quad \cplP \underset{\mathrm{cosmo}}{<} 8 \times 10^{-2}.
\end{equation}

In the following, we are interested in the observable consequences the
previously discussed boundary inflation eras and their resulting
primordial power spectra may have on the CMB.  The fact that the
$\mat$ field decays into radiation, matter and dark matter implies
that there are no observable entropy modes between the produced
cosmological fluids. Of course they are present between the
cosmological fluids and the moduli fields $\modR$ and $\modP$, however
we will assume that any back reaction effects on the minimally coupled
fluids can be neglected as soon as the theory does not deviate too
much from General Relativity after inflation. As a result, only the
adiabatic and tensor primordial power spectra derived in
Sect.~\ref{sect:linpert} source the observed cosmological
perturbations. The current cosmological data being sensitive to the
tilt and amplitude of the primordial power
spectra~\cite{Tegmark:2002cy, Peiris:2003ff, Leach:2003us,
  Barger:2003ym, Bond:2004rt, Leach:2005av}, one may expect to probe
the model parameters involved. In particular, since for $\cplP=0$ our
model matches with single field chaotic inflation, this parameter will
be used to analyse whether the boundary inflation model is favored or
excluded by the data compared to a single field model.

Keeping in mind that the present model has to verify
Eq.~(\ref{eq:firstgradtoday}), we will nevertheless only use a weak
late-time cosmology upper bound $\cplP<0.2$ as a prior in the
following CMB computations~\cite{Nagata:2003qn,
  Acquaviva:2004ti}. Indeed, previous derivations of the CMB
anisotropies in scalar-tensor gravity theory have been focused on the
late-time modifications the non-minimally coupled scalar fields may
produce by assuming standard power law primordial power spectra. The
present analysis being precisely concerned with the primordial stages
and the generation of the cosmological perturbations, it may give a
complementary view of the effect expected on the CMB in scalar-tensor
theories.

\begin{figure}
\begin{center}
\includegraphics[width=8.5cm]{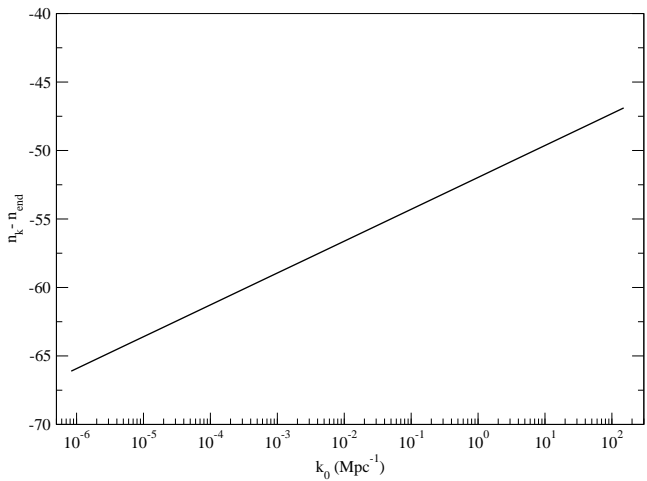}
\caption{Correspondence between the wavenumber today $\kmpc$
  ($\Mpc^{-1}$) and the bfold of Hubble exit. The background model is
  the same as in Fig.~\ref{fig:powerspec}.}
\label{fig:bfoldtok}
\end{center}
\end{figure}

Under these assumptions, one can approximate $\kappaeff \equiv \kappa
\Afac \simeq \kappacav$ and Eq.~(\ref{eq:redshiftinf}) can be further
simplified into~\cite{Liddle:2003as}
\begin{equation}
\label{eq:aendinf}
\ln\dfrac{\aJF_\uend}{\aJF_\zero}  \simeq \dfrac{1}{2} \ln\left(
  \sqrt{2 \OmegaJF_\urad} \,  \kappaeff \hubbleJF_\zero \right) -
\dfrac{1}{4} \ln \dfrac{2 \kappaeff^4 \energyJF_\uend}{3},
\end{equation}
where $\aJF_\zero$, $\OmegaJF_\urad$ and $\hubbleJF_\zero$ are,
respectively, the scale factor, the total density parameter of
radiation and the Hubble parameter today~\cite{Liddle:1993fq}. Note
that $\kappaeff^4 \energyJF_\uend \simeq \kappa^4 \energy_\uend$ in
the Einstein frame for small variations of the conformal factor during
the late-time cosmology. {}From Eq.~(\ref{eq:aendinf}), we can relate
the wavenumber of the observed perturbations today to the
corresponding comoving wavenumber during their primordial generation
\begin{equation}
  \label{eq:kcosmo}
  \dfrac{\kwav}{\calH} = \constconv \,
  \dfrac{\kmpc}{\hubblend} \, \dfrac{\aJF_\zero}{\aJF_\uend}
  \, \ue^{-(\bfold)},
\end{equation}
where $\kmpc$ is the comoving wavenumber today in units of $\Mpc^{-1}$,
$\hubblend = \kappa \hubble$ and $\constconv$ a conversion unit
constant $\constconv \simeq \exp(-130)$. In Fig.~\ref{fig:bfoldtok} we
have plotted the correspondence between $\kmpc$ and the bfold of
Hubble exit during inflation for the same model parameters as in
Fig.~\ref{fig:powerspec}, and with $\aJF_\zero=1$.

\subsection{CMB power spectra}

In order to compute the multipole moments of the CMB anisotropies we
have used a modified version of the $\camb$
code~\cite{Lewis:1999bs}. {}From Sect.~\ref{sect:background} and
Sect.~\ref{sect:linpert}, in addition to the usual cosmological
parameters, the CMB anisotropies in the boundary inflation model under
scrutiny are characterized by $\Afac_\uini$, $\modR_\uini$, and
$\cplP$ for the non-minimally coupled sector as well as $\matnd_\uini$
and $\baremassnd$ in the matter sector [see Eqs.~(\ref{eq:branesetup})
  and (\ref{eq:potmat})]. However the absolute value of the conformal
factor $\Afac$ is non-observable and can be arbitrarily set at a given
time, \eg $\Afac_\uini=1$. As a result, it is more convenient to
parametrize the boundary inflation eras in terms of the rescaled
parameters: $\Afac_\uini \matnd_\uini$, $\baremasseff \equiv
\Afac_\uend \baremassnd = \kappaeff \baremass$, with $\modR_\uini$ and
$\cplP$ unchanged. For given values of these four parameters the
background solution is computed and used to setup the cosmological
framework by means of Eqs.~(\ref{eq:aendinf}) and (\ref{eq:kcosmo}).
Then, each observable wavenumber required in the primordial adiabatic
and tensor power spectra is computed from its corresponding quantum
initial state as presented in Sect.~\ref{sect:linpert}, and thereby
used to derive the resulting CMB power spectra. According to the
discussion in Sect.~\ref{sect:linpert_power}, we do not expect
observable effect on the CMB coming from $\Afac_\uini \matnd_\uini$
and $\modR_\uini$ since for a wide range of values they do not modify
the primordial power spectra. However, these parameters have been kept
free in the following in order not to reject the rare cases for which
the end of the $\modR$-dominated expansion era would precisely occur
during the generation of the largest observable perturbation modes.
\begin{figure}
\begin{center}
\includegraphics[width=8.5cm]{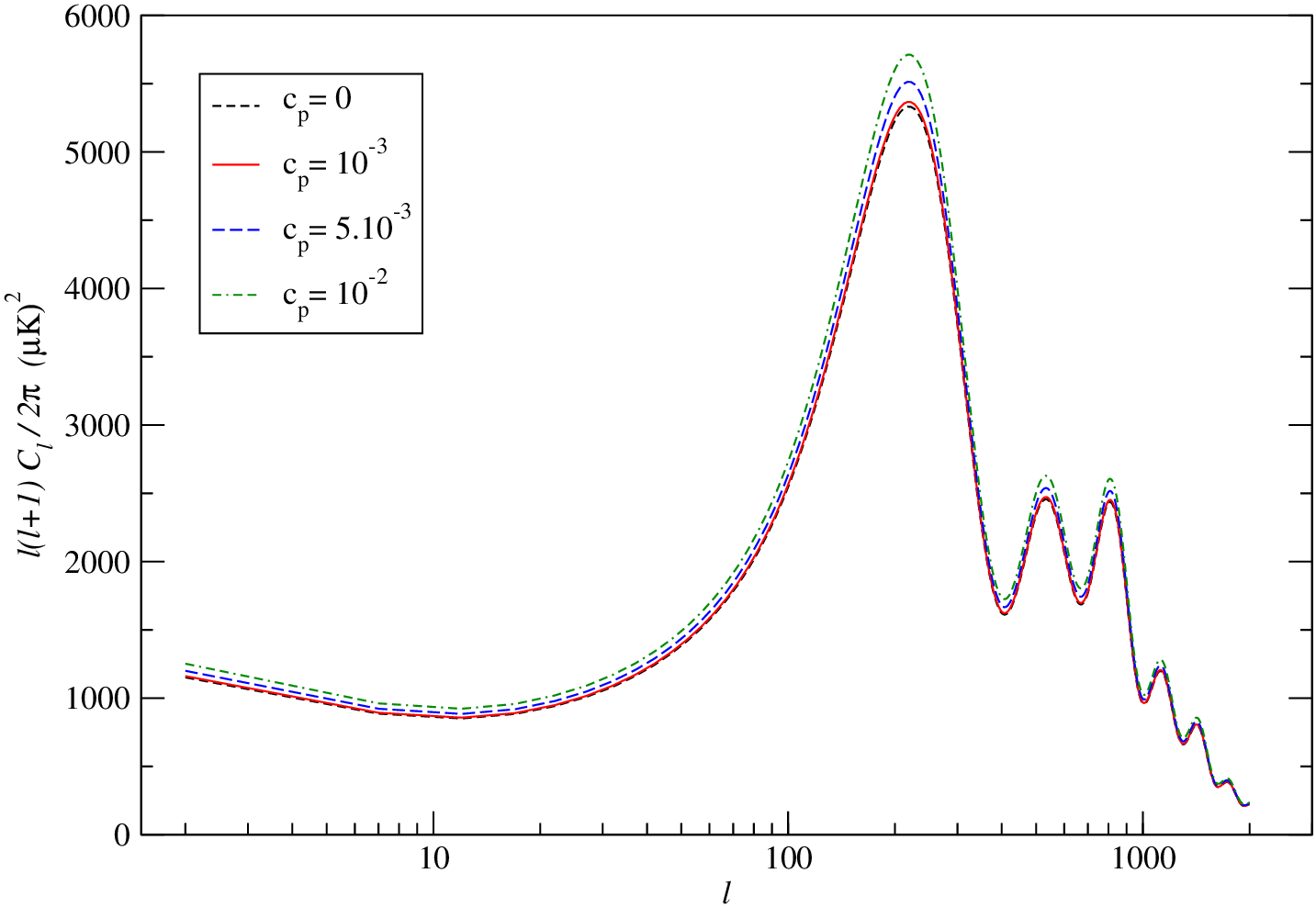}
\caption{Dependence of the $TT$ power spectra with respect to the
  moduli coupling constant $\cplP$, at fixed effective matter field
  mass $\baremasseff \equiv \Afac_\uend \baremassnd$.}
\label{fig:cmbevol}
\end{center}
\end{figure}
On the other hand, Fig.~\ref{fig:cmbevol} confirms that varying the
moduli coupling constant $\cplP$ is not innocuous for the temperature
angular power spectrum. However, as previously mentioned, the
effective matter field mass $\baremasseff$ is also directly involved
in the amplitude of the primordial power spectra through
Eq.~(\ref{eq:hubblesquare}) and one may expect a degeneracy between
these parameters on their respective CMB influence. In addition, both
$\cplP$ and $\baremasseff$ have been shown to modify the tilts of the
primordial power spectra in Sect.~\ref{sect:linpert_power}.

In the next section, a Monte Carlo Markov Chain (MCMC) exploration of
the cosmological and primordial parameter space if performed by using
the first year WMAP data. Note that the method we are using relies on
a full numerical scheme which does not involve any approximation of
the primordial power spectra.

\subsection{MCMC exploration}

Following the method of Ref.~\cite{Lewis:2002ah}, we consider a
parameter space involving a minimal set of cosmological parameters:
the ratio of the sound horizon to the angular diameter distance
$\thetalewis$ (related to the reduced Hubble parameter $h$), the
density parameter of baryons $\Omegab$ and cold dark matter $\Omegac$,
the optical depth $\optdepth$ (or the redshift of reionization
$\zre$); as well as our primordial parameters: $\Afac_\uini
\matnd_\uini$, $\modR_\uini$, $\cplP$ and $\baremasseff$. Note that
the four primordial parameters fix the overall and relative amplitudes
of the primordial scalar and tensor power spectra. The MCMC
computations have been done by using the $\cosmomc$
code~\cite{Lewis:2002ah} calling the modified $\camb$ version based on
our inflationary code, and given the first year WMAP data and the
associated likelihood code~\cite{Kogut:2003et,
  Verde:2003ey,Hinshaw:2003ex}.

In order to check the relevance of our full numerical
approach, we have first performed a MCMC exploration on the single
field chaotic inflation model which is obtained by fixing
$\modR_\uini=\cplP=0$. Indeed, the current constraints on the
cosmological parameters using the WMAP data usually assume either
power law or slow-roll approximated primordial power
spectra~\cite{Peiris:2003ff,Barger:2003ym,Leach:2003us}. Since our
method goes further it may be interesting to check its consistency
with the current existing bounds. Moreover, the single field chaotic
model will be our reference model to discuss the more complex
features associated with the boundary inflation model.

\subsubsection{Chaotic model}

\begin{figure}
\begin{center}
\includegraphics[width=8.5cm]{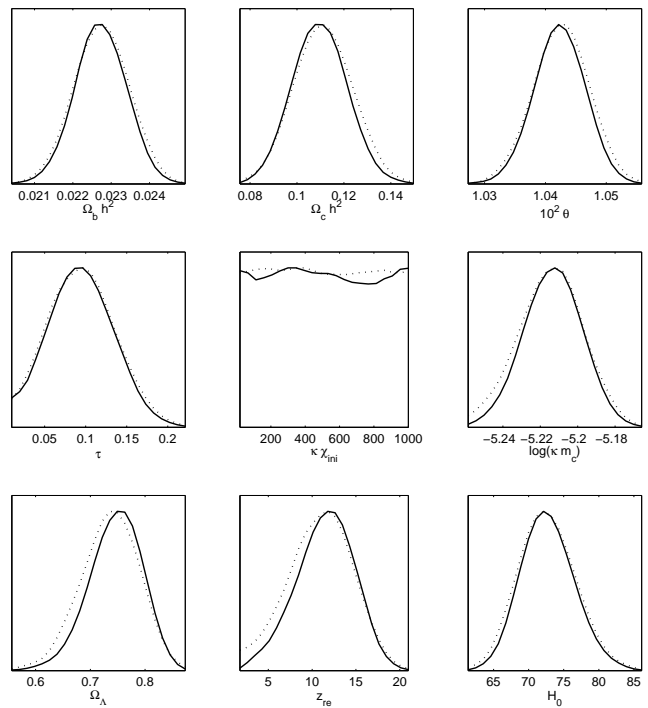}
\caption{The 1D marginalized posterior probability distributions in
  the pure chaotic inflation model. The dotted lines are the
  associated 1D mean likelihoods.}
\label{fig:chaotic_1D}
\end{center}
\end{figure}

\begin{figure}
\begin{center}
\includegraphics[width=8.5cm]{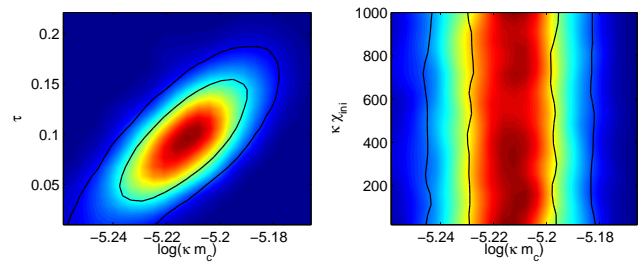}
\caption{$1\sigma$ and $2\sigma$ isocontours of some correlated 2D
  marginalized posterior distributions in the pure chaotic inflation
  model. The colormap traces the 2D mean likelihoods. The inflaton
  mass fixes the amplitude of the primordial perturbations and is
  constrained around Grand Unification values.}
\label{fig:chaotic_2D}
\end{center}
\end{figure}

We have used standard prior distributions for the base cosmological
parameters (see Ref.~\cite{Lewis:2002ah}) whereas wide top hat priors
have been chosen for the chaotic model parameters: $20 < \matnd_\uini
< 1000$, and $-10 < \log(\baremassnd) < 0$. The lower limit on the
initial matter field value is set in order to get the right order or
magnitude of the minimal total number of efolds required to solve the
flatness and homogeneity problem ($\simeq 60$). In addition, in the
$\cosmomc$ code, we have coded a ``hard prior'' rejecting any model
for which observable perturbations cannot be initially set in a
Bunch-Davies vacuum with $\constdec=100$: for the chaotic model this
can occur when there is not enough efolds of inflation [see
  Eq.~(\ref{eq:initime})]. The higher limit has been chosen in order
to avoid prohibitive computation time which occurs when the total
number of efolding become very large. The prior on $\log(\baremassnd)$
is chosen to contain the value required to get the right amplitude of
the CMB anisotropies. The obtained posterior probability distributions
for the base and derived cosmological, as well as the primordial
parameters are plotted in Fig.~\ref{fig:chaotic_1D} and
Fig.~\ref{fig:chaotic_2D}. They correspond to $50000$ samples for
which the errors on their shape do not exceed $3\%$. Note that they
do not rely on any slow-roll approximation but only on the linear
perturbation theory and the cosmological setup of
Sect.~\ref{sect:cmb_cosmoframe}.


Firstly, the constraints on the base cosmological parameters are found
to be consistent with the current state of the art~\cite{Spergel:2003cb,
  Peiris:2003ff, Leach:2003us, Tegmark:2003ud}. For the chaotic model
parameters, the inflaton mass $\baremass$ is well constrained, as
expected from a parameter involved in the normalization of the
primordial power spectra, and we find, at $2\sigma$ level
\begin{equation}
\label{eq:2sigbaremass}
-5.24 <\log(\kappa \baremass) < -5.18.
\end{equation}

As expected, the initial value of the field is found to be
unconstrained. Concerning the overall ability of the chaotic
model to fit the data, the best fit is obtained with a likelihood of
$-\ln(\like) \simeq 714.4$ for $1343$ degrees of freedom which
render the chaotic model slightly more favored than its slow-roll
approximated version.

In the next section, a MCMC exploration is performed for the boundary
inflation model. Since in the limit $\cplP\rightarrow0$ this model
matches with the chaotic one, the posterior probability distribution
of the coupling constant $\cplP$ traces how favored or disfavored the
boundary inflation model is with respect to the single field chaotic
model.

\subsubsection{Boundary inflation model}

\begin{figure}
\begin{center}
\includegraphics[width=8.5cm]{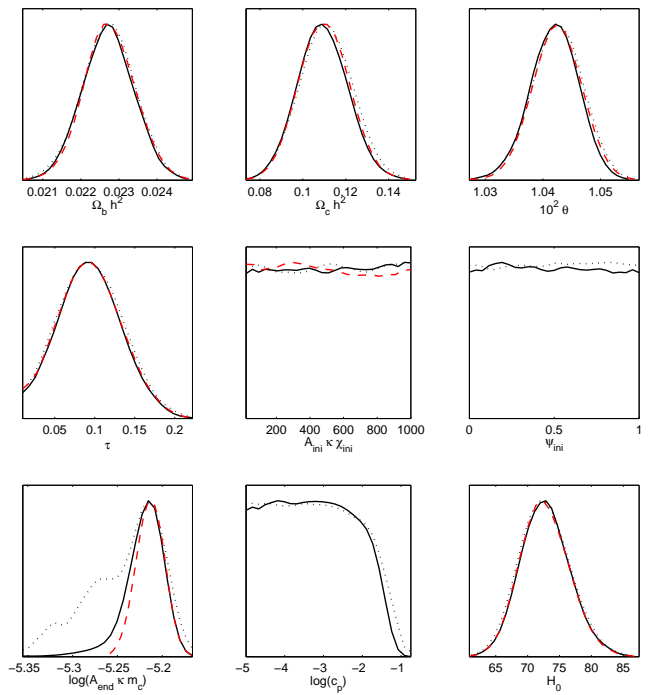}
\caption{The 1D marginalized posterior probability distributions
  (black solid lines) and the 1D mean likelihoods (dotted lines) in
  the boundary inflation model. Only the base cosmological and
  primordial parameters are represented, together with the chaotic
  model posteriors of Fig.~\ref{fig:chaotic_1D} (red dashed line).
  These posteriors have been derived under the prior choice
  $\log(\cplP)<0.2$ for which the moduli do not strongly influence the
  late-time cosmology.}
\label{fig:scaltens_1D}
\end{center}
\end{figure}

\begin{figure}
\begin{center}
\includegraphics[width=8.5cm]{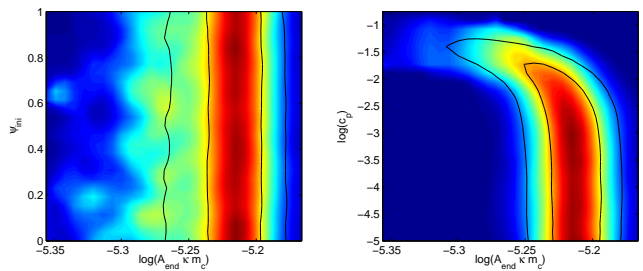}
\caption{$1\sigma$ and $2\sigma$ isocontours of the correlated 2D
  marginalized posterior distributions in the boundary inflation
  model. High values of the coupling constant $\cplP$ lead to bigger
  tilt in the primordial power spectra and are strongly cut. The
  degeneracy between the effective mass of the matter field and the
  moduli coupling constant is seen on the right plot.}
\label{fig:scaltens_2D}
\end{center}
\end{figure}

The full set of primordial parameters is now considered: $\Afac_\uini
\matnd_\uini$, $\modR_\uini$, $\cplP$ and $\baremasseff$.  Compared to
the chaotic model, there are two additional parameters $\modR_\uini$
and $\cplP$, and a rescaled one: $\baremasseff \equiv \Afac_\uend
\kappa \baremass$. Remember that the latter no longer refers to the
bare mass of the matter field but to its effective mass at the end of
inflation. A flat prior has been chosen for the moduli coupling
constant: $-10 < \log(\cplP) < -0.75$. The lower limit corresponds to
the decoupling between the $\modP$ and $\matnd$ fields for which the
boundary model is equivalent to single field chaotic inflation,
whereas the upper limit correspond to the weak post-inflationary bound
above which the moduli influence the late-time cosmology (see
Sect.~\ref{sect:cmb_cosmoframe}). Concerning the $\modR_\uini$ values,
we have set a conservative top hat prior $0 < \modR_\uini < 1$ since,
according to our $\constdec=100$ prescription, larger values of
$\modR$ cannot be observed from the CMB point of view (see
Sect.~\ref{sect:linpert_numres}). The other priors for the base
cosmological parameters and the effective mass are the same as for the
chaotic model. Their corresponding posterior distributions have been
obtained for $160000$ samples and are plotted in
Fig.~\ref{fig:scaltens_1D} and Fig.~\ref{fig:scaltens_2D} where the
marginalized distributions obtained for the chaotic model are also
superimposed (red dashed lines).


On one hand, the robustness of the cosmological parameters with
respect to the primordial power spectra is recovered: there are no
significant deviations between their posterior probability
distributions in the chaotic model and in the boundary inflation
model.

On the other hand, we did not find a better fit of the WMAP data with
the boundary inflation primordial power spectra. More precisely, the
best fit has the same likelihood than the one obtained with the
chaotic model: $-\ln(\like)=714.4$ but now for $1341$ degrees of
freedom. Statistically, the boundary inflation model is disfavored by
its additional two parameters which do not help to improve the
fit~\cite{Lewis:2002ah, Lazarides:2004we, Trotta:2005ar}. As expected
from the previous discussions, the initial values of the background
fields $\Afac_\uini \matnd_\uini$ and $\modR_\uini$ have not
observable effect and are consequently unconstrained by the
data. Furthermore, the marginalised probability distribution of the
coupling constant $\cplP$ in Fig.~\ref{fig:scaltens_1D} exhibits a
plateau for $\cplP \lesssim 10^{-3}$. Since in the decoupling limit
$\cplP \rightarrow 0$ the boundary model identifies with the single
field chaotic model, the existence of such a plateau shows that the
WMAP data are not sensitive to the deviation induced by the moduli
during inflation as long as $\cplP<10^{-3}$. As a result, from the
data point of view, all these inflationary models, boundary and
chaotic, cannot being distinguished in that regime.

However, the rapid decay of the $\cplP$ marginalised probability
distribution for values larger than $10^{-2}$ signals the sensitivity
of the data in this regime, which disfavour the corresponding boundary
inflation models. More precisely, one obtains the two-sigma level
upper bound for the moduli coupling constant
\begin{equation}
\log(\cplP) < -1.64 \,,
\end{equation}
which can be recast to an upper limit for the scalar-tensor conformal
gradient
\begin{equation}
\label{eq:firstgradpower}
\alphafac^2 \equiv \metric^{ab} \alphafac_{a} \alphafac_{b} < 6 \times 10^{-4},
\end{equation}
at $95\%$ confidence level. In terms of the post-Newtonian parameters
more commonly used to constrain scalar-tensor
theories~\cite{Will:2001mx}, we have
\begin{equation}
1-\gamma = \dfrac{4 \alphafac^2}{1 + 2\alphafac^2} < 2 \times 10^{-3}.
\end{equation}

In fact, this constraint comes from the effects discussed in
Sect.~\ref{sect:linpert_power}: the tilt of the primordial power
spectra in the boundary model increases with the value of the the
moduli coupling constant due to the running of the effective matter
field mass. Since the WMAP data are compatible with flat primordial
power spectra, it is therefore not surprising that they end up
constraining deviations from scale-invariance. This effect also
explains why the marginalised probability distribution of the
effective mass $\baremasseff=\Afac_\uend \baremassnd$ is skewed to
lower values. For a given value of $\baremasseff$, larger values of
$\cplP$ implies a more efficient running of the conformal factor
$\Afac$ during inflation, and in particular larger values of $\Afac
\baremassnd$ at the time when the observable perturbations have left
the Hubble radius (see Fig.~\ref{fig:powerevol} and
Fig.~\ref{fig:compare}). This degeneracy between $\cplP$ and
$\baremasseff$ is clearly seen in Fig.~\ref{fig:scaltens_2D}: the
two-dimensional marginal probability is distorted in such a way that
the overall amplitude of the primordial perturbations fit with the one
measured by WMAP.

Let us emphasize again that Eq.~(\ref{eq:firstgradpower}) holds at the
time of inflation and comes only from the shape of the primordial
power spectra. In fact, for a given matter sector, this bound
certainly applies to any inflationary scalar-tensor models involving a
running conformal factor during the generation of primordial
perturbations. On the contrary, scalar-tensor models exhibiting a
relaxation mechanism toward General Relativity are expected to evade
this bound: as it is the case for the moduli $\modR$ in our setup,
such scalar gravity fields would reach the minimum of the potential
after only a few efolds of inflation thereby freezing the running of
the conformal factor. In any case, comparing
Eq.~(\ref{eq:firstgradpower}) to the current solar-system and
astrophysical bounds in Eqs.~(\ref{eq:firstgradtoday}) and
(\ref{eq:firstgradcosmo}) shows that scalar-tensor inflation should be
considered in the search of scalar-tensor effects in cosmology.

\section{Conclusion}

In this paper, we have studied the CMB signatures a minimal realistic
boundary inflation model may have in a (reasonably) restricted
parameter space in which the observable effects come only from the
shapes of the primordial scalar and tensor power spectra.

Firstly, it has been shown that the background evolution involves
three different eras corresponding to the domination of one of the
three fields over the others. According to the values of the coupling
constants, these eras may or may not be of inflationary kind. In our
case, one of the moduli field, namely $\modR$, is associated to a
non-accelerated expansion during which it rapidly relaxes toward
vanishing values. Once $\modR$ trapped into its minimum, two smoothly
connected inflationary periods driven by the moduli $\modP$ and the
matter field $\mat$ take place until the matter field oscillates
around the minimum of its potential thereby starting a reheating
period.

At the perturbation level, we have discussed the generation and
observability of the primordial cosmological perturbations arising
during these inflationary eras. Since the $\modR$-dominated era
precedes the inflationary eras, this field has not significant
observable effects. Moreover, the existence of an attractor in the
background fields evolution during inflation erases any memory of the
initial conditions when one is concerned with the observable
perturbations. As a result, only the effective mass $\baremasseff$ of
the matter field and the moduli coupling constant $\cplP$ end up being
of observable interest for the CMB. The former encodes the amplitude
of the primordial perturbations at Hubble exit and the latter
quantifies the changing rate of the conformal factor with respect to
the evolution of the moduli $\modP$. Moreover, as expected for a
multifield system, there are entropy perturbations which source the
adiabatic modes after Hubble exit. However, for cosmologically
relevant values of the moduli coupling constant $\cplP$, the entropy
and adiabatic modes are found to be weakly coupled and the adiabatic
power spectrum dominates at the end of inflation. Nevertheless, even
in this regime, the running of the conformal factor at Hubble exit has
been shown to significantly increase the tilt of the adiabatic power
spectrum.

We have then assumed a toy cosmological framework in order to compute
the seeded CMB anisotropies. In this framework, the moduli $\modR$ and
$\modP$ survive in the late-time cosmology whereas the standard
cosmological fluids are produced by the decay of the matter field
$\matnd$. In order to study the observable effects stemming from
inflation only, we have focused on values of the moduli coupling
constant which do not strongly modify the late-time cosmology, namely
$\cplP<0.2$. Under this prior choice, we have computed the induced CMB
anisotropies and performed a MCMC analysis of the parameter space
given the first year WMAP data. The boundary inflation model appears
to be indistinguishable from a single field chaotic model as long as
$\cplP < 10^{-3}$ whereas it is disfavored for larger values of the
coupling constant. Since the overall best fit model lies in the $\cplP
< 10^{-3}$ region of the parameter space, the boundary inflation model
is statistically disfavored compared to a single chaotic model by its
two additional degrees of freedom which do not help to improve the
fit. Moreover, the current WMAP data lead to a $95\%$ marginalized
probability bound $\log(\cplP) < -1.64$ which corresponds to the
post-Newtonian Eddington parameter upper limit $1-\gamma < 2 \times
10^{-3}$.

The above bound is not competitive with the solar-system upper limit
and remains only slightly stronger than the late-time cosmological
one. However, it holds at the time of inflation and provides in this
respect a very early constraint to the scalar-tensor models which
would behave during inflation as the boundary model, but would relax
toward General Relativity afterwards. A word of caution is in order
since this constraint certainly does not apply to all scalar-tensor
inflationary models.  Indeed, it essentially relies on the running of
the conformal factor during the generation of the primordial
perturbations, and also depends on the shape of the matter
potential. For example, one may imagine to freeze the running of the
conformal factor by stabilising the moduli in their bulk potential
[see Eq.~(\ref{eq:actionbrane})]. However, in view of the future more
accurate data, this work suggests that a fully consistent derivation
of scalar-tensor theory bounds in the context of cosmology should
involve both an inflationary and post-inflationary modelisation of the
expected deviations from General Relativity.

The present work has been focused on the inflationary eras only and
for simplicity we have not considered reheating phenomena that might
modify the evolution of the cosmological perturbations. It could be
interesting in future works to quantify these effects in view of the
CMB data. In addition, we have not considered the post-inflationary
scalar-tensor effects. A natural extension of the current work would
be to perform a MCMC exploration on the moduli coupling constants by
using simultaneously scalar-tensor inflationary and post-inflationary
codes to compute the cosmological perturbations. In particular, one
could quantify the late-time effects of the entropy perturbations
existing generically between the moduli and the cosmological fluids in
such brane world scenarios.

\acknowledgments

We thank Nicole Audiffren, Carlo Contaldi, Gilles Esposito-Far\`ese,
Samuel Leach, J\'er\^ome Martin, Christophe Rhodes, Roberto Trotta and
Jean-Philippe Uzan for enlightening discussions and advices on the
different parts of this work. The computations have been performed for
one part in the Centre Informatique National de l'Enseignement
Sup\'erieur~\cite{cines} during one CPU-time year, and for the other
thanks to a substantial time allocation by the French data processing
center for Planck-HFI~\cite{planck} and by the U.~K. Computational
Cosmology Consortium~\cite{cosmos}. This work is supported in part by
PPARC.

\bibliography{bibboundinf}

\end{document}